\documentclass[sigconf]{acmart}
\newcommand{\quotes}[1]{\textit{``#1''}}

\usepackage{enumitem}
\usepackage{subcaption}

\usepackage{listings}
\newcommand{\revision}[1]{\textcolor{black}{#1}}

\AtBeginDocument{%
  }

\copyrightyear{2026}
\acmYear{2026}
\setcopyright{cc}
\setcctype{by}
\acmConference[CHI '26]{Proceedings of the 2026 CHI Conference on Human Factors in Computing Systems}{April 13--17, 2026}{Barcelona, Spain}
\acmBooktitle{Proceedings of the 2026 CHI Conference on Human Factors in Computing Systems (CHI '26), April 13--17, 2026, Barcelona, Spain}
\acmDOI{10.1145/3772318.3790379}
\acmISBN{979-8-4007-2278-3/2026/04}

\begin{document}

\title[How Avatar Appearances Shape Perceptions of AI Hiring]{Skin-Deep Bias: How Avatar Appearances Shape Perceptions of AI Hiring}
\thanks{This is the author's version of the work. It is posted here for your personal use. Not for redistribution. The definitive Version of Record was published in \textit{Proceedings of the 2026 CHI Conference on Human Factors in Computing Systems (CHI '26)}. \url{https://doi.org/10.1145/3772318.3790379}.}

\author{Ka Hei Carrie Lau}
\email{carrie.lau@tum.de}
\affiliation{%
\department{Chair of Human-Centered
Technologies for Learning}
\department{Munich Center for Machine Learning (MCML)}
\institution{Technical University of Munich}
\city{Munich}
\country{Germany}}

\author{Philipp Stark}
\email{philipp.stark@keg.lu.se}
\affiliation{%
\institution{Department of Human Geography, Lund University}
\city{Lund}
\country{Sweden}}

\author{Efe Bozkir}
\email{efe.bozkir@tum.de}
\affiliation{%
\institution{Chair of Human-Centered
Technologies for Learning, Technical University of Munich}
\city{Munich}
\country{Germany}}

\author{Enkelejda Kasneci}
\email{enkelejda.kasneci@tum.de}
\affiliation{%
\institution{Chair of Human-Centered
Technologies for Learning, Technical University of Munich}
\city{Munich}
\country{Germany}}

\renewcommand{\shortauthors}{Lau et al.}

\begin{abstract}

Artificial intelligence is increasingly used in hiring, raising concerns about how applicants perceive these systems. While prior work on algorithmic fairness has emphasized technical bias mitigation, little is known about how avatar identity cues influence applicants’ justice attributions in an interview context. We conducted a crowdsourcing study with 215 participants who completed an interview with photorealistic AI avatars varied in phenotypic traits (race and sex), followed by a standardized rejection. Using self-reports, sentiment analysis, and eye tracking, we measured perceptions of trust, fairness, and bias. Results show that racial mismatch heightened perceptions of ethnic bias, while partial match (sharing only one identity) reduced fairness judgments compared to both full and no match. This work extends the Computers-Are-Social-Actors paradigm by demonstrating that avatar appearances shape justice-related evaluations of AI. We contribute to HCI by revealing how identity cues influence fairness attributions and offer actionable insights for designing equitable AI interview systems.

\end{abstract}

\ccsdesc[500]{Human-centered computing~Empirical studies in HCI}
\ccsdesc[500]{Human-centered computing~Empirical studies in collaborative and social computing}

\keywords{generative AI, crowdsourcing, social identity, fairness}

\begin{teaserfigure}
  \includegraphics[width=\textwidth]{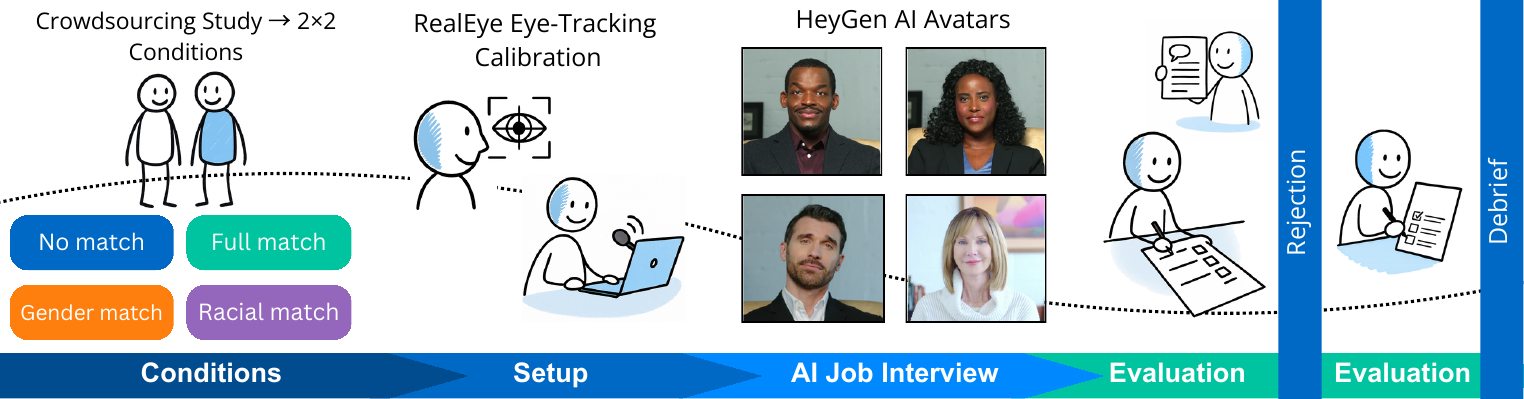}
  \caption{Study overview. We recruited participants via crowdsourcing and assigned them to a 2×2 experimental design with four avatar–participant match conditions (No Match, Gender Match, Racial Match, Full Match). The experiment procedure included eye-tracking calibration, a real-time verbal AI interview, post-interview measures, a scripted rejection, post-outcome measures, and a debrief. We created some illustrative icons with GPT-5 and then the author edited them to align with the study design and visual style.}
  \Description{Flow diagram of the study procedure and 2-by-2 avatar--participant matching design. Participants are recruited online and assigned to one of four conditions: no match, gender match, racial match, or full match. The timeline shows eye-tracking calibration, a spoken AI interview, a scripted job rejection message, post-interview ratings, post-outcome evaluations, and debrief.}
  \label{fig:teaser}
\end{teaserfigure}

\maketitle

\section{Introduction}

Artificial intelligence (AI) systems are increasingly deployed in high-stakes contexts such as college admissions or hiring interviews~\cite{Busum2023, Salvetti2024, Chipotle2024}. Many of these tools use embodied conversational agents (ECAs) with human-like voices and appearances to simulate face-to-face interaction. Equipped with large language models (LLMs) as their backend, ECAs can now respond adaptively to users in real time.  Their efficiency and scalability have accelerated adoption in recruitment, where companies seek faster and more standardized processes~\cite{Cowgill2019bias}. AI hiring platforms are often promoted as objective and fair, with claims that they can reduce or even eliminate human bias in evaluation~\cite{HireVue2025, Neufast2025}. However, it remains unclear how applicants perceive issues of fairness and trust when engaging with these systems.

This uncertainty around users’ perceptions points to an already established paradox in human–computer interaction (HCI). While realistic interfaces are designed to foster trust by making interactions feel natural, in high-stakes contexts, this same realism can amplify disappointment and perceptions of unfairness when outcomes are unfavorable. Prior research has examined this paradox from two perspectives. On the one hand, \citet{Wang2020} show that fairness perceptions depend not only on group-level bias but also on whether individuals personally benefit or lose from an algorithmic decision. On the other hand, \citet{Nass2000} describe human interaction with realistic interfaces as “mindless social responses”: people do not thoughtfully anthropomorphize machines but instead apply social scripts automatically. These perspectives suggest that fairness perceptions are shaped by both the outcomes of algorithmic decisions and automatic, unreflective social responses. In this work, we focus on the latter. These reactions become more consequential when lifelike avatars clearly display social category cues, such as race and gender, which can lead people to automatically apply social stereotypes to them~\cite{Nass1994, Reeves1996mediaeq}. If these systems behave in ways that echo offline discrimination, users can interpret this as algorithmic unfairness and a breach of trust~\cite{Woodruff2018}. We extend this line of research to embodied contexts, where avatars’ phenotypic traits, such as race and sex, can act as salient social cues. We use the term~\textbf{phenotypic bias transfer} to describe how avatars’ \textbf{phenotypic traits} can trigger social categorization processes that reproduce racialized and gendered stereotypes during interaction. These phenotypic traits are visible features such as skin color and secondary sex characteristics (e.g., facial structure, jawline, hairstyle). In this paper, we use \emph{sex} to refer to avatars’ phenotypic appearance (female/male) and \emph{gender} to refer to participants’ self-identified category (woman/man). Accordingly, we use the term \textbf{gender match} to denote when an avatar’s sex aligns with the participant’s self-identified gender, and \textbf{racial match} to denote when an avatar’s race aligns with the participant’s self-identified ethnicity.

Investigating these potential perceptual biases is urgent, since conversational and generative artificial intelligence (Gen-AI) ECAs are already being rapidly adopted in socially impactful domains such as customer service and recruitment~\cite{GKINKO2023102568, Sacar2024, Szafarski2025}. Yet research still lags behind, particularly in understanding how people perceive fairness and trust in these systems and how, beyond questions of algorithmic fairness, they may harm racially and culturally marginalized groups~\cite{Schlesinger2017, Hankerson2016}. Beyond this societal urgency, our study also offers a theoretical contribution. Social Identity Theory (SIT) predicts that people show in-group favoritism, evaluating those who share their identity more positively~\cite{Tajfel1974}, while the Computers Are Social Actors (CASA) paradigm shows that people apply human social rules to machines~\cite{Nass1994}. As a result, these perspectives suggest that AI systems can elicit social behaviors comparable to those observed in human–human interaction, which in turn may introduce biases. It is therefore important to examine how this unfolds in simulated high-pressure contexts such as AI hiring, where adaptive ECAs are used and the outcome is unfavorable. Methodologically, most previous studies rely on scripted or pre-recorded agents, limiting adaptive and interactive capabilities, and to the best of our knowledge, no prior work has examined real-time Gen-AI-based ECAs in the context of simulated job interviews.

Building on these considerations, our study analyzes perceptions of trust, fairness, and bias along the phenotypic categories of race and sex, and examines the interaction between these two categories. We treat these effects as structural rather than incidental, in line with Schlesinger et al.~\cite{Schlesinger2017}, who argue that treating identity categories in isolation overlooks how technologies reinforce overlapping systems of oppression. Furthermore, interactional designs can reproduce existing social hierarchies, and Zajko~\cite{Zajko2022} shows that algorithmic systems embed and reproduce broader structural inequalities. Together, these perspectives highlight that perceived fairness problems in AI extend beyond algorithmic justice. Systems may satisfy procedural fairness criteria yet still create unequal experiences through avatar~\textbf{phenotypic traits} that trigger biased responses. \revision{Closest to our work, Biswas et al.~\cite{Biswas2024} varied avatar race and gender in pre-recorded asynchronous video interviews (AVI); we extend this line of work by examining perceived fairness as a relational judgment in a real-time interview setting.} Specifically, we ask the following research questions (RQs):

\begin{enumerate}
  \item[\textbf{RQ1.}] Do participants perceive meaningful differences between avatar phenotypic traits?
  \item[\textbf{RQ2.}] Does racial or gender matching affect trust, perceived fairness, and bias?
  \item[\textbf{RQ3.}] Does racial or gender matching affect implicit behavioral measures (sentiment and eye tracking)?
\end{enumerate}

To investigate this problem, we developed an online platform that simulates real-time~\textbf{AI interviews} using a Gen-AI-ECA powered by the HeyGen avatar generator. In a crowdsourcing study with 215 participants, we employed a $2\times2$ between-subjects design that manipulated avatar~\textbf{phenotypic traits} (race: black/white; sex: male/female) to either match or mismatch participants’ own identities. Each participant completed the interview verbally and received a standardized rejection outcome, providing a context that approximates applied AI-mediated interviews, while controlling for~\emph{outcome favorability} bias shown by~\citet{Wang2020}. We collected both self-report measures (fairness, trust, bias) and implicit behavioral data (sentiment, webcam-based eye tracking), which provide a complementary perspective on how participants engaged with the AI interviewer.~\revision{Our findings show that post-interview trust was high across conditions, while perceptions of fairness and bias shifted.} Racial mismatches increased perceived bias and attention to the avatar’s face, while partial matches lowered perceived fairness.

Our work makes three contributions to HCI and AI fairness research. We follow the taxonomy of research contributions in HCI~\cite{Wobbrock2016} and build on recent discussions of LLMs-related contributions at CHI~\cite{Pang2025}, contributing a methodological advance alongside theoretical and empirical insights.

\begin{description}

\item [Theoretical]  
We show that race–gender identity cues do not operate separately; partial matches reduced perceived fairness compared to both full matches and mismatches. This intersectional fairness paradox challenges SIT predictions and extends CASA by showing how fairness attributions emerge when multiple social cues interact in adaptive AI interviews.

\item [Empirical]  
We provide experimental evidence that fairness and bias perceptions are more sensitive to avatar identity cues than trust.~\revision{Post-interview trust was high across conditions}, while racial mismatches heightened perceived bias and partial matches reduced perceived fairness, highlighting that identity cues in adaptive ECAs can shape user experience in consequential settings such as AI-hiring.

\item [Methodological]  
We developed a scalable platform for studying ECA interactions using real-time generative avatars with multimodal measurement. By combining self-report, sentiment analysis, and webcam-based eye tracking, the platform provides a reusable framework for capturing both explicit perceptions and implicit behaviors in fairness-sensitive applications.

\end{description}

\section{Related Work}
We review and contrast our work with prior research in three areas that frame our study: (1) SIT in human–AI interaction, (2) social response in ECAs, and (3) fairness in AI hiring.

\subsection{Social Identity Theory and AI}

Social Identity Theory (SIT), introduced by Tajfel~\cite{Tajfel1974}, explains how individuals categorize themselves and others into social groups, producing in-group favoritism and out-group bias~\cite{Hogg2006}. Tajfel’s minimal group experiments showed that such favoritism can emerge even from arbitrary categories~\cite{Tajfel1970}. We argue that these mechanisms are equally relevant to human–AI interaction. When AI systems present social cues such as faces, voices, or names, they can trigger the same categorical thinking observed in human groups. This echoes Zajko’s~\cite{Zajko2022} call to examine not only algorithmic bias, but also how AI systems and their interfaces participate in reproducing structural inequalities beyond technical fairness metrics.

Recent research has begun to examine SIT in human–AI interaction. Seaborn~\cite{Seaborn2025} provides a theoretical foundation, outlining axioms that show how anthropomorphic cues can trigger categorization processes while emphasizing that such processes remain \emph{dynamic} (shifting during an interaction) and \emph{context-dependent} (activated by situational factors). Sun et al.~\cite{sun2024} demonstrate that users place greater trust in AI agents that are both ingroup and humanoid, extending similarity–attraction effects to artificial agents. Edwards et al.~\cite{edwards2019evaluations} extend this line of work by showing that students high in age identification rated an older-sounding AI instructor as more credible, motivating, and socially present, suggesting that role stereotypes (e.g., professors as older) can outweigh simple similarity cues. In interview contexts, identity effects appear even less predictable. Biswas et al.~\cite{Biswas2024} found that participants’ own demographics, mediated by Social Presence and Perception (SPP), shaped perceived fairness whereas avatar race and gender showed no main effects. Do et al.~\cite{DoTiffany2024} reported that mismatched avatars in virtual reality (VR) reduced embodiment and disproportionately harmed minority users. Together, these findings indicate that SIT effects in human–AI interaction vary with \textbf{identity salience} and the \textbf{stakes of interaction}.

However, prior work has not examined how multiple identity cues intersect in consequential settings. \citet{Schlesinger2017} call for intersectional approaches, warning that treating identity categories in isolation overlooks how technologies reproduce overlapping systems of oppression. Our study addresses these gaps by testing \emph{intersectional identity cues} (race $\times$ gender) in a simulated AI interview and assessing how they influence perceived fairness.

\subsection{Social Response in ECAs}

\subsubsection{From Minimal Cues to LLM-Driven ECAs}
Embodied conversational agents (ECAs) convey verbal and non-verbal cues (e.g., gaze, gesture) that render abstract AI processes into socially recognizable identities~\cite{BAILLY2010598, Kang2015}. Early CASA research showed that even minimal cues, such as synthetic voice, role, or gendered presentation, could elicit politeness and stereotyping in simple, scripted systems~\cite{Nass1994}. Building on this, \citet{Reeves1996mediaeq} argued that people respond to media as if it were real because human cognition cannot reliably distinguish mediated from unmediated social cues. Later studies confirmed this claim: as computer agents became more anthropomorphic, they elicited stronger social judgments, social influence, and trust than abstract forms~\cite{GONG20081494}. However, today’s LLM-based ECAs combine human-like cues such as natural dialogue, turn-taking, and photorealistic appearance. This creates a dilemma for HCI: the same cues designed to foster natural interaction may also trigger stereotyping in consequential domains. As ECAs become more realistic and socially responsive, they are shifting from low-stakes applications to high-stakes contexts where decisions can carry real-life consequences for individuals. Yet most prior work has examined ECAs in scripted or low-stakes settings, leaving open the question of whether existing theories still hold when adaptiveness and stakes increase. To address this gap, our study situates ECAs in a simulated job hiring context and combines self-reports with implicit measures to capture perceptions. Table~\ref{tab:related_work} positions our study within this trajectory by comparing prior ECA research on identity cues with our real-time, AI-based interview, multimodal approach.

\begin{table*}[t]
\centering

\caption{Trajectory of related work on ECAs and identity cues. Prior studies are mostly low-stakes, scripted, and based on self-reports. Our study introduces a real-time, simulated AI-based job interview and multimodal testbed approach.}
\Description{Table summarizing prior work on embodied conversational agents and identity cues. Rows list studies and their context and methodology; columns compare setting stakes, level of scripting, interaction mode, identity cue manipulation, measures used (self-report versus behavioral), and key findings. The last row highlights the present work as a real-time simulated job interview with multimodal measures.}
\label{tab:related_work}

\begin{tabular}{@{}
p{0.14\linewidth}
p{0.22\linewidth}
p{0.19\linewidth}
p{0.37\linewidth}
@{}}
\toprule
\textbf{Study} & \textbf{Context (Domain stakes)} & \textbf{Interface} & \textbf{Key contribution} \\
\midrule

Nass et al.~\cite{Nass1994}
& Lab quiz, Q\&A tasks (low)
& Scripted agent
& CASA paradigm: minimal cues (voice, role, gender) elicit social responses \\

Sun et al.~\cite{sun2024}
& Trust game, lab \& online (low)
& Video intro \& game interface
& Ingroup and humanoid AIs increase trust \\

Aumüller et al.~\cite{Aumuller2024}
& Bank service chatbot (low)
& Text chat and avatar (ambiguous vs.\ abstract)
& Participants preferred abstract icons; perceiving the avatar as female increased intention to use \\

Szafarski et al.~\cite{Szafarski2025}
& Luxury automotive customer interaction (medium)
& Video demo (Study~1); live LLM-based ECA (Study~2)
& High TAM acceptance, with younger users rated more intuitive; design and user experience focus; no fairness analysis \\

Biswas et al.~\cite{Biswas2024}
& Simulated job interview (high)
& Asynchronous video interview (AVI)
& No main effect of agent race/gender; participant demographics shaped fairness via SPP; no match-mismatch; no post\mbox{-}outcome analysis \\

\midrule
\textbf{This study}
& Simulated job interview (high)
& Real-time~\mbox{LLM-based} ECA
& First perceived fairness evaluation of participant--avatar identity alignment with multimodal measures (self-report, sentiment, eye tracking) \\
\bottomrule
\end{tabular}
\end{table*}

\subsubsection{Beyond Self-Report to Multimodal Measures}

Most ECA studies still rely on self-reports, overlooking how subconscious processes shape interaction~\cite{Biswas2024, sun2024, Aumuller2024, Szafarski2025, Seaborn2025}. \citet{Reeves1996mediaeq} argue that self-reports are insufficient because human responses to media are often unconscious and automatic, making self-reflections unreliable. Similarly, \citet{Grimm2010} shows that self-reports are vulnerable to \textit{social desirability bias}, which can mask implicit prejudice. Together, these limitations make it particularly difficult to evaluate perceptions of socially sensitive topics.

Recent work has begun to address this gap by using implicit behavioral measures to capture processes that self-reports miss. Eye-tracking research has long shown that gaze patterns reveal cultural and social dynamics. In the other-race effect (ORE), observers allocate attention differently to own- versus other-race faces, which contributes to poorer cross-race recognition~\cite{HILLS2013586}. In live interaction, cultural groups also differ in how often they engage in mutual gaze~\cite{Haensel2022}. More recent work connects gaze to attitudes. \citet{Steinfeld2021} found that Israeli participants who looked longer at a Palestinian speaker during simulated virtual contact reported more positive outgroup perceptions. Extending beyond gaze, \citet{Peck2021} showed that head and hand motion in VR revealed racial bias even when measures such as shooting accuracy and response latency did not.

These findings demonstrate that multimodal implicit measures uncover attentional, cultural, and bias-related processes that explicit judgments often suppress or overlook. More broadly, \citet{Lai2023} caution that much empirical research on human--AI decision making relies on simplified tasks and ad hoc measures, raising concerns about validity and generalizability. While our study remains a simulation, it moves closer to being ecologically valid by situating ECAs in an online hiring interview and combining validated self-reports with implicit behavioral measures. Building on this, we contribute a scalable multimodal platform for evaluating both explicit and implicit perceptions in fairness-sensitive ECA interactions.

\subsection{Fairness in AI Hiring}

Before AI was introduced, fairness in hiring was studied extensively in organizational psychology. Organizational justice theory shows that applicants judge fairness by both outcomes and procedures such as job-relatedness, consistency, and respectful treatment~\cite{Gilliland1993, Colquitt2001}. Building on this, Ryan and Ployhart~\cite{Ryan2000} show that applicants’ experiences of the hiring process influence their trust, willingness to pursue opportunities, and perceptions of organizational legitimacy.

With the rise of automated screening and AI-based decision systems~\cite{Li2021, Soni2024}, fairness concerns have become more critical. While such tools promise efficiency and scalability, they also reduce human oversight, introduce opacity, and risk embedding or amplifying inequities. Much of the research on AI hiring has therefore emphasized bias mitigation and transparency through technical approaches~\cite{Albaroudi2024}. Experimental evidence further shows that awareness of algorithmic gender bias can deter qualified women from applying~\cite{Edwin2025}. Yet this line of research still frames fairness primarily as an algorithmic property rather than as a lived experience. \citet{Liao2022} argue that fairness is shaped not only by algorithms but also by how users interpret model explanations, while \citet{Woodruff2018} emphasize that such perceptions are especially salient for minoritized groups, who evaluate fairness through the lens of their lived experiences. Van Berkel et al.~\cite{Niels2019} further demonstrate that people actively judge which predictors feel fair or unfair, underscoring that fairness is ultimately constructed through perception rather than technical definitions alone. Our study aligns with this line of work by examining how perceived fairness is shaped in a simulated AI interview, particularly through identity cues expressed by avatars.

Recent work also foregrounds fairness as an experiential judgment, examining it through system qualities and social cues. Hosain et al.~\cite{HOSAIN2025} show that applicants’ perceptions of AI hiring systems are positively associated with procedural justice, mediated by whether the systems feel easy to use, useful, and trustworthy. This highlights that fairness judgments extend beyond technical bias reduction to include applicants’ broader impressions of system usability. At the same time, the design of AI interview interfaces matters. Biswas et al.~\cite{Biswas2024} varied avatar race and gender in an AVI-based study and treated agent and participant demographics as independent predictors. They found no main effects of avatar race or gender, but provided an important foundation for our work by showing that participants’ own demographics, mediated by SPP, can influence perceived fairness, privacy, and impression management. Our work extends this research by studying perceived fairness in a different setting, where applicants must interpret a uniformly negative decision, and by analyzing relational identity (mis)matches between participants and LLM-based interviewers using multimodal behavioral measures. Beyond the avatar interface, other communicative cues can also shape applicants’ experiences. Heo et al.~\cite{Heo2025} found that gendered chatbot voices affected applicants’ language use, confidence, and interviewer ratings. Pyle et al.~\cite{Pyle2024} further showed that candidates experienced emotion AI in video interviews as procedurally and interactionally unjust, raising concerns about its adoption.

Despite these advances, research on fairness in AI hiring remains narrow. Most studies emphasize algorithmic design or rely on self-reports of fairness, offering limited insight into how judgments develop and shift during live interaction. Our study addresses this gap by examining how fairness perceptions unfold in real-time exchanges with photorealistic ECAs, and by analyzing how unfavorable outcomes are attributed to the system. By combining explicit justice ratings with implicit measures of attention and sentiment, we advance a multimodal approach for examining fairness in adaptive ECA hiring contexts.

\section{Method}

In this section, we describe the experimental design, demographics of our participants, experimental platform, procedure, measures, and analysis. The Institutional Review Board (IRB) of the Technical University of Munich approved our study.

\subsection{Experimental Design}

We designed a $2 \times 2$ between-subjects experiment to investigate how identity matching between participants and an AI interviewer in terms of race and sex influences user perceptions of a simulated job interview. To this end, we randomly assigned our participants to one of four experimental conditions:

\begin{itemize}
    \item \textbf{Avatar race:} Match vs.\ Mismatch (relative to the participant’s self-identified ethnicity)
    \item \textbf{Avatar sex:} Match vs.\ Mismatch (relative to the participant’s self-identified gender)
\end{itemize}

\begin{figure}[t]
\centering
\setlength{\fboxsep}{0pt}

\begin{minipage}[t]{0.49\columnwidth}
  \centering
  \fbox{\includegraphics[width=\linewidth]{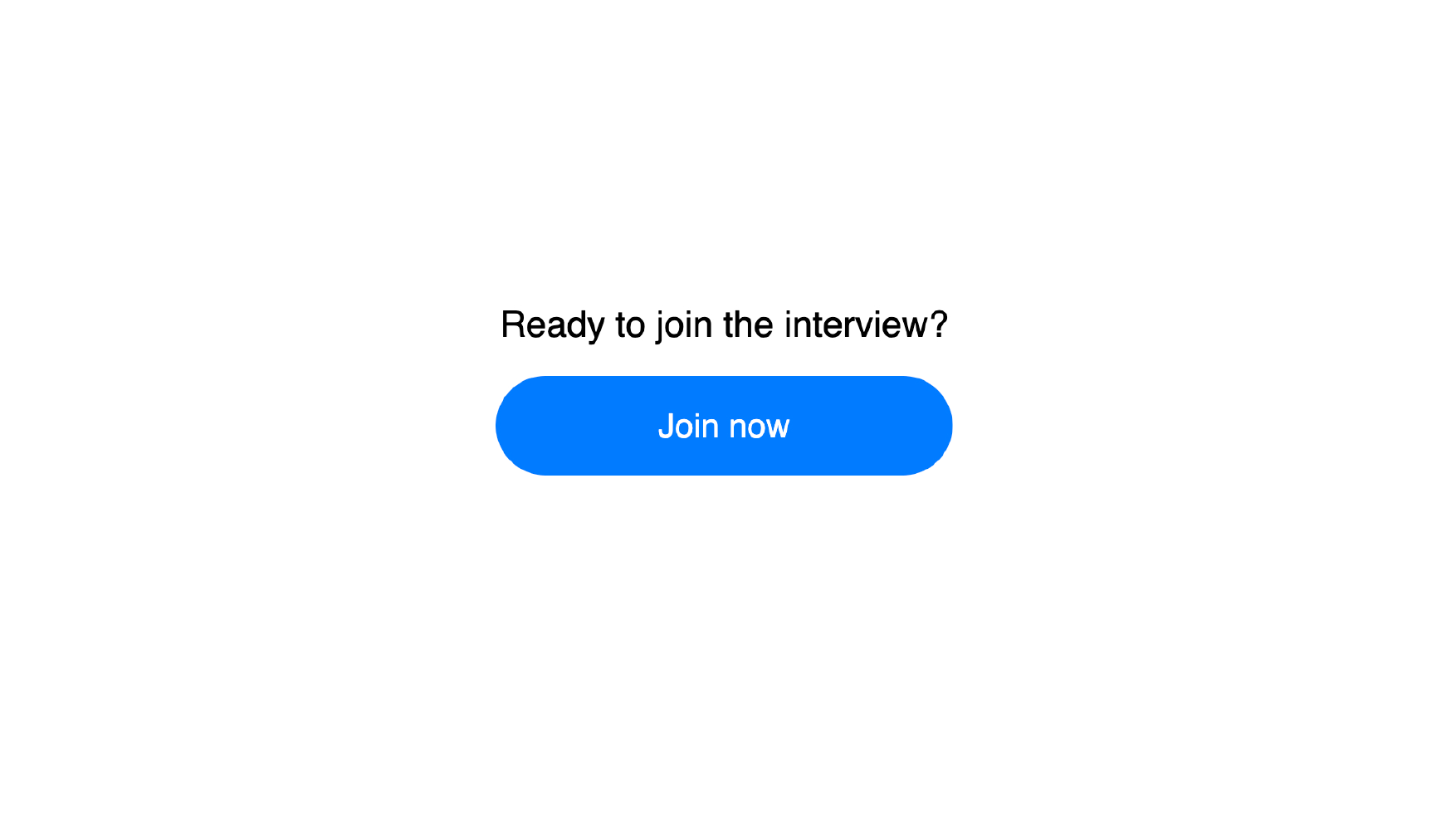}}
  \caption*{\textbf{(a)} Join screen}
\end{minipage}\hfill
\begin{minipage}[t]{0.49\columnwidth}
  \centering
  \fbox{\includegraphics[width=\linewidth]{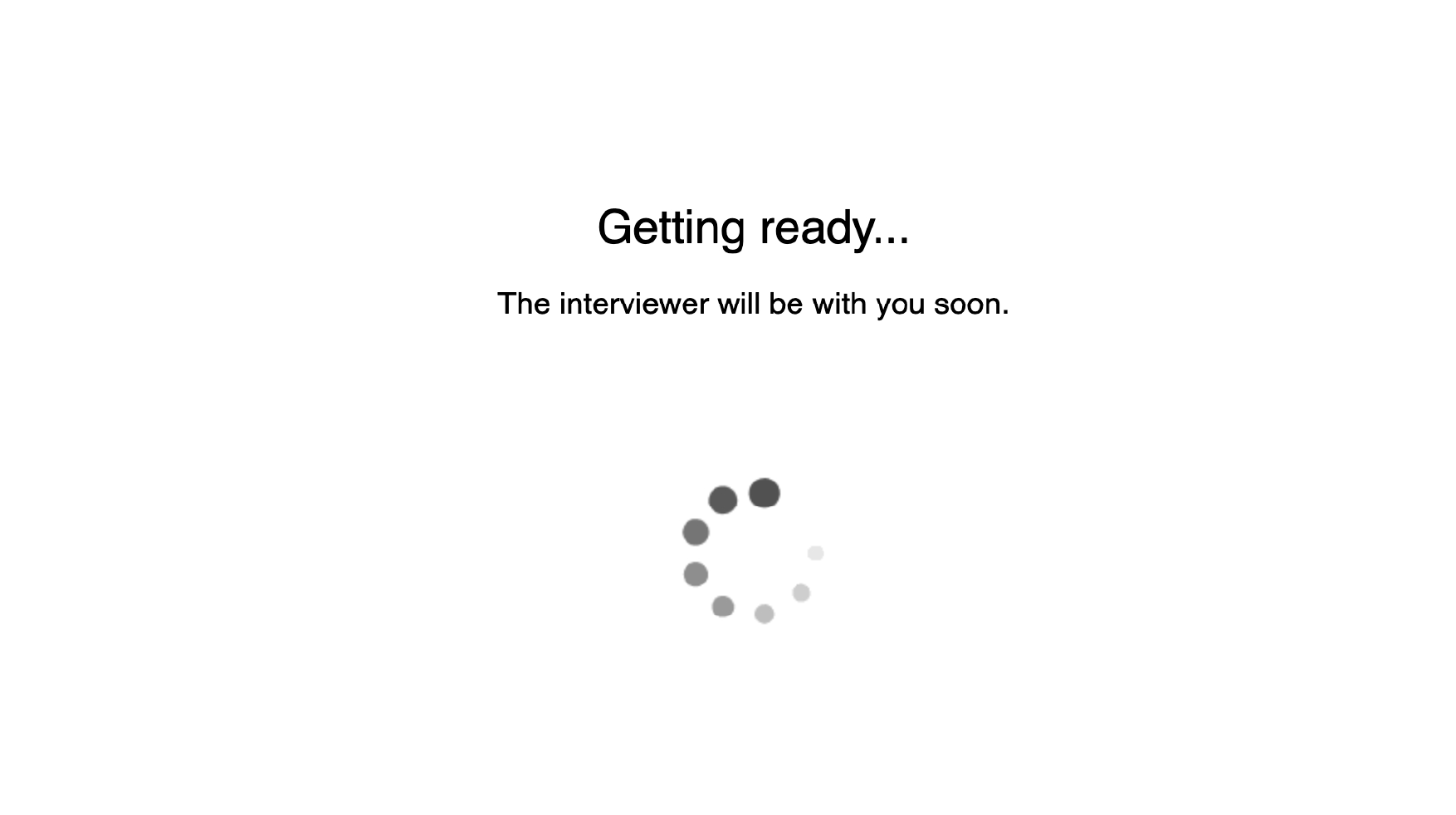}}
  \caption*{\textbf{(b)} Waiting room}
\end{minipage}

\begin{minipage}[t]{0.85\columnwidth}
  \centering
  \fbox{\includegraphics[width=\linewidth]{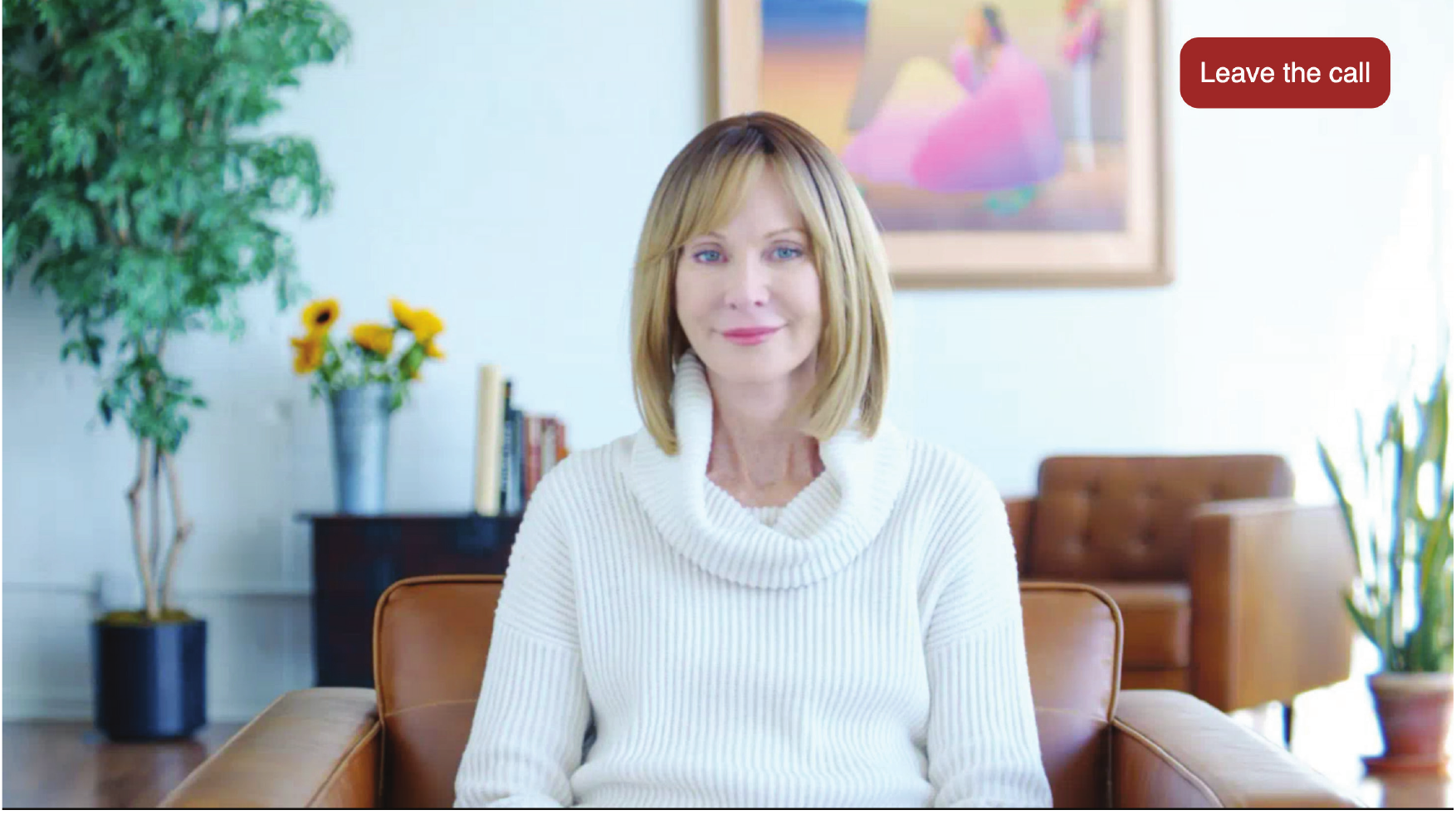}}
  \caption*{\textbf{(c)} In-call view}
\end{minipage}

\caption{Participants accessed the study through a video-call-like interface with a join screen, waiting room, and in-call view (with a “Leave the call” button) designed to mirror familiar online meeting tools.}
\Description{Participants accessed the study through a video-call-like interface with a join screen, waiting room, and in-call view (with a “Leave the call” button) designed to mirror familiar online meeting tools.}
\label{fig:video_interface}
\end{figure}

Each participant completed the study in a single session, which involved answering four voice-based interview questions that the AI interviewer delivered. To make the experience feel realistic, we utilized a user interface that resembles a typical online interview platform. Participants joined the session through a video-call-like interface, with buttons and layout elements that mirrored the appearance of online meeting tools, as shown in Figure~\ref{fig:video_interface}. In contrast to prior AVI work~\cite{Biswas2024}, our AI interviewer responded in real time using a streaming LLM-based ECA, enabling turn-taking, clarifications, and conversational adaptivity.

\subsubsection{Interview Task Design}\label{sec:interview-task-design}
We designed our interview and feedback procedure based on previous research in human resources and organizational psychology~\cite{raisova2012comparison, Zibarras2015, Campion1997, Levashina2014}. In this study, we informed our participants that they would be interviewed for a ``customer support'' role. We selected this role for its relatability and low task complexity, especially in a remote, online setting with a diverse participant pool. Furthermore, customer support positions emphasize interpersonal communication, empathy, and problem-solving competencies, which are linked to job readiness~\cite{Lindsay2015}. 

To assess these competencies, we followed a Competency-Based Interview (CBI) format, a structured technique that focuses on assessing specific competencies relevant to the targeted job, rather than relying on past experiences as in Behavioral Event Interviews (BEI). CBI formats are considered fairer and, in some cases, more predictive of job performance~\cite{raisova2012comparison, Zibarras2015}. The interview questions were designed in a way that progressed from general to role-specific, including: 
\textit{(1) Could you tell me a bit about yourself?, (2) Can you explain a situation where you helped another person and solved a problem for them?, (3) How do you respond to critical feedback?, and (4) How would you handle a frustrated customer?}

All participants, regardless of condition, received a simulated standardized rejection. This approach aligns with~\citet{Skitka2003}, who distinguish outcome favorability from outcome fairness, and is further supported by Gilliland’s fairness model~\cite{Gilliland1993}, which emphasizes that applicant reactions are shaped not only by outcomes but also by procedural qualities (i.e., consistency, bias suppression, and respectful treatment). These factors are especially influential when outcomes are unfavorable, helping us assess whether identity matching moderates negative perceptions, which were not examined in prior AVI studies~\cite{Biswas2024}. The rejection message followed best practices outlined by~\citet{Cortini2019}, using polite, informal language that acknowledged participants’ effort without over-justifying the decision to deliver the message. The rejection message used for this study is provided in Appendix~\ref{appendix:rejection_message}.

\subsection{Participants}

We conducted the study in July 2025 on the online research platform Prolific\footnote{\url{https://www.prolific.com/}, last accessed 22 January 2026}, recruiting 228 participants from the United Kingdom, Germany, and the United States. Eligibility criteria included being at least 18 years old, fluent in English, and having access to a functional webcam, microphone, stable internet connection, and a quiet environment. For our $2 \times 2$ factorial design, we restricted inclusion to participants who, according to Prolific’s prescreen settings\footnote{\url{https://researcher-help.prolific.com/en/article/412c0a}, last accessed 22 January 2026}, identified their sex as female or male and their ethnicity (simplified) as white or black. In our own survey, participants reported self-identified gender and ethnicity again using expanded categories (e.g., non-binary, mixed ethnicity). For analysis, we included only those who self-identified as women or men and as either white (European descent) or black (African descent); these self-reports were then used to assign match/mismatch conditions relative to avatars’ race and sex.

We focused on this majority–minority contrast given evidence that hiring discrimination is often triggered by visible identity cues such as skin color, which signal perceived cultural distance~\cite{Zschirnt2016}, and that black applicants receive significantly fewer callbacks than equally qualified white applicants~\cite{Bertrand-Mullainathan}.

We randomly assigned participants to one of four conditions, crossing avatar race (white vs.\ black) and sex (female vs.\ male). We used an adaptive stratified allocation method (minimization) to balance participant numbers across conditions, with randomization applied to break ties when multiple cells were equally under-represented. Gender was recorded as women ($n=110$), men ($n=108$), and non-binary ($n=2$). Although our design required binary assignment to avatar conditions, we report the presence of two non-binary participants for completeness; they, along with others whose self-reported identity did not fit the predefined categories (e.g., mixed ethnicity), were excluded from analysis.

We excluded thirteen participants (5.7\%), either due to technical issues (e.g., avatar not loading, eye-tracking errors, incomplete sessions) or because their reported identity fell outside our $2 \times 2$ design categories. The final analysis sample comprised 215 participants. Table \ref{tab:demographics} summarizes participant demographics by experimental condition. Each session lasted approximately 20 minutes, and participants were compensated \pounds4.27 (equivalent to \pounds12.81/hour) in line with Prolific’s compensation policy\footnote{\url{https://researcher-help.prolific.com/en/articles/445230-prolific-s-payment-principles}, last accessed 22 January 2026}. Compensation was not based on interview performance or the hiring decision, and we did not offer any bonuses. All participants provided informed consent and were reminded of their right to withdraw at any time.

\begin{table*}[tb!]
\centering
\caption{Participant demographics by experimental condition. Values are mean $\pm$ SD unless otherwise noted.}
\Description{Table of participant demographics and background measures by experimental condition (no match, gender match, racial match, full match). Columns show sample sizes; rows include counts for gender, ethnicity, and vision status, percentages for employment and interview experience, and mean ± SD for age, AI usage, speaking nervousness, and fairness beliefs/attitudes, with footnotes defining categories.}
\label{tab:demographics}
\begin{tabular}{lcccc}
\toprule
 & \multicolumn{4}{c}{Experimental Condition} \\
\cmidrule(lr){2-5}
Variable & No match & Gender match & Racial match & Full match \\
 & (n=53) & (n=57) & (n=54) & (n=51) \\
\midrule
\textbf{Demographics} & & & & \\
~~Gender, n (Men/Women)\textsuperscript{a} & 27/26 & 27/30 & 28/26 & 26/25 \\
~~Age (years) & 38.5 $\pm$ 12.7 & 39.3 $\pm$ 11.0 & 39.3 $\pm$ 9.5 & 42.5 $\pm$ 12.6 \\
~~Ethnicity, n (White/Black)\textsuperscript{b} & 27/26 & 29/28 & 29/25 & 28/23 \\
~~Vision status, n (Normal/Glasses/Contacts)\textsuperscript{c} & 37/13/3 & 36/20/1 & 38/12/4 & 33/16/2 \\
\midrule
\textbf{Interview Experience} & & & & \\
~~Employment, \% employed\textsuperscript{d} & 75 & 79 & 81 & 78 \\
~~Oral interview experience, \% moderate+\textsuperscript{d} & 68 & 81 & 71 & 76 \\
\midrule
\textbf{Individual Differences} & & & & \\
~~AI usage (hours/week) & 6.0 $\pm$ 12.9 & 7.9 $\pm$ 12.2 & 4.6 $\pm$ 7.2 & 4.5 $\pm$ 4.8 \\
~~Speaking nervousness (1--10 scale) & 4.2 $\pm$ 2.3 & 4.0 $\pm$ 2.1 & 4.3 $\pm$ 2.4 & 4.1 $\pm$ 2.2 \\
~~Fairness beliefs and attitudes (1--5 scale) & 3.22 $\pm$ 0.35 & 3.29 $\pm$ 0.35 & 3.29 $\pm$ 0.34 & 3.14 $\pm$ 0.34 \\
\bottomrule
\multicolumn{5}{l}{\footnotesize\textsuperscript{a}Two participants identified as non-binary overall.}\\
\multicolumn{5}{l}{\footnotesize\textsuperscript{b} Ethnicity categories reflect survey wording: White (for example European descent), Black or African descent.}\\
\multicolumn{5}{l}{\footnotesize\textsuperscript{c} Vision status = self-reported normal vision, wearing glasses, or wearing contact lenses.}\\
\multicolumn{5}{l}{\footnotesize\textsuperscript{d} Employment = full/part-time; Oral interview experience = moderate or higher.}\\
\end{tabular}
\end{table*}

\subsection{Experimental Platform}
\begin{figure}[b]
    \centering
    \includegraphics[width=\columnwidth]{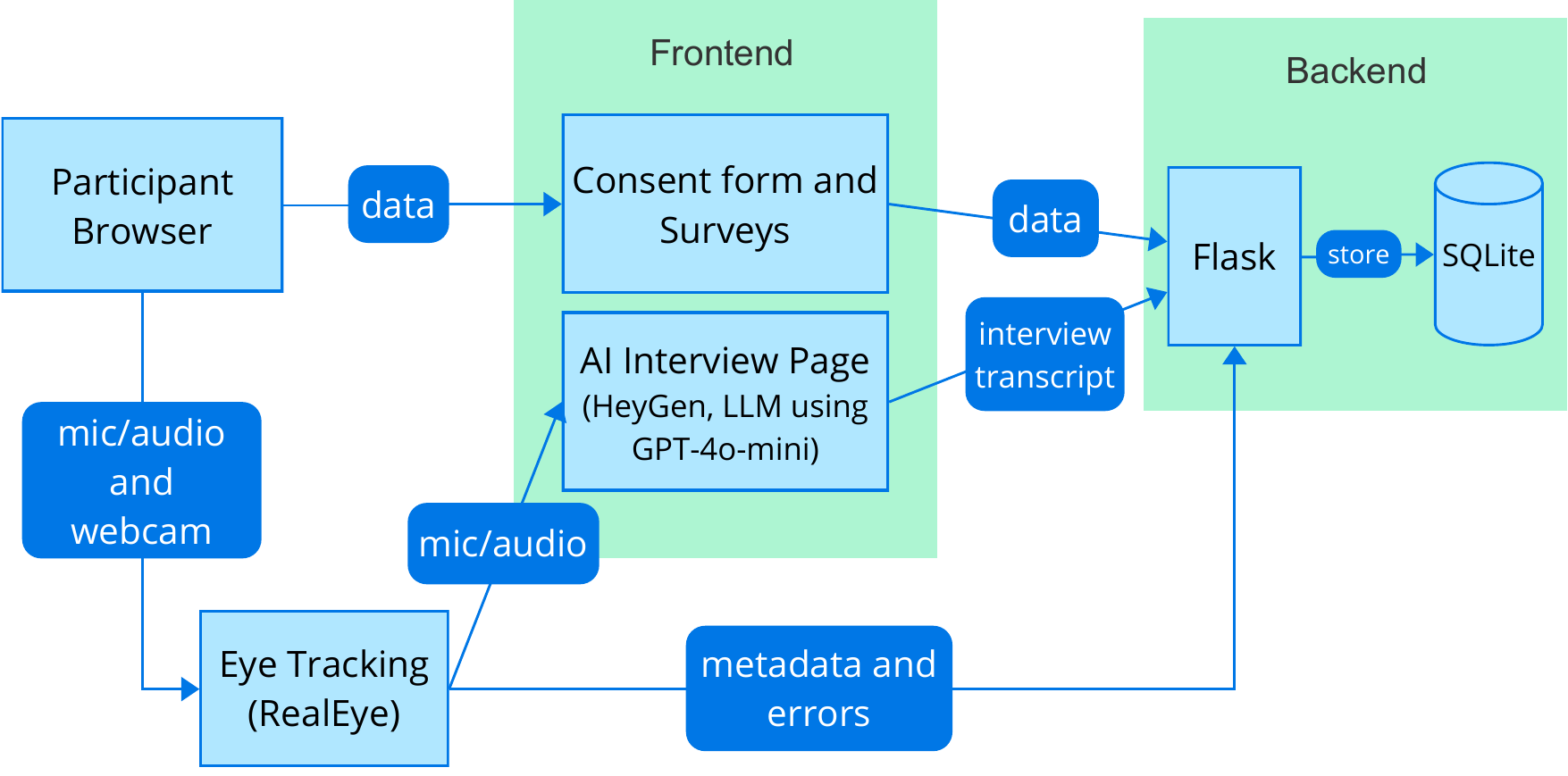}
    \caption{High-level system architecture of the experimental platform.}
    \Description{Block diagram of the experimental platform. A participant connects via a web browser (frontend) to complete a consent form and surveys, and to interact with an AI interviewer. The interviewer runs through HeyGen AI avatar. Eye-tracking data (from the participant’s device) is collected alongside interaction data. The frontend sends data via an API and a local SQLite database to a backend service built with Flask.}
    \label{fig:system_flow}
\end{figure}

We developed a web-based platform that managed the survey flow, the AI interviewer, and webcam-based eye tracking. The overall system architecture is shown in Figure~\ref{fig:system_flow}. The platform was hosted on a remote server and accessed via standard web browsers (except Mozilla Firefox, due to API limitations). Participants joined remotely using their own laptops or desktops with a stable internet connection, webcam, and microphone.

\subsubsection{System Architecture}
We implemented the backend in Flask version 3.1.1\footnote{\url{https://flask.palletsprojects.com/en/stable/}, last accessed 22 January 2026} and the frontend in vanilla JavaScript with Vite, controlling the experimental workflow. The platform delivered the informed consent form digitally, enabled simultaneous capture of questionnaire responses and interview transcripts via speech-to-text (STT), and integrated RealEye for eye-tracking data collection.

\subsubsection{AI Interviewer and Conversational AI}
We built the AI interviewer with HeyGen's Streaming Avatar SDK\footnote{\url{https://docs.heygen.com/docs/streaming-api}, last accessed 22 January 2026}, which renders photorealistic avatars in real time with synchronized lip movements and facial expressions. We selected four professionally dressed avatars from HeyGen's library, systematically varying in phenotypic features associated with race (skin tone, facial structure) and sex (jawline, hairstyle) while maintaining consistent professional attire and setting. These visual traits were not intended to represent any individual's subjective identity but to evoke perceived social categories commonly used in social cognition research. The four avatars used in the experiment are shown in Figure~\ref{fig:teaser}, and all shared the same neutral background and the same voice settings (American English, neutral pitch and pace, calm tone).

We generated avatar responses through HeyGen’s integration with OpenAI’s GPT-4o-mini model\footnote{\url{https://platform.openai.com/docs/guides/realtime}, last accessed 22 January 2026}, which supports low-latency and real-time voice conversation. In our experiment, the avatar could greet participants by name, acknowledge information they had provided, ask for clarification when an answer was unclear, and pose brief follow-up questions when responses were very short. This real-time setup differs from prior AVI systems~\cite{Biswas2024}, where the virtual interviewer presented pre-recorded video prompts rather than generating adaptive follow-ups in real time. These adaptive behaviors can make the interaction feel more natural and conversational, thereby improving the ecological validity of the simulated AI-based job interview.

We controlled the interviewer’s behavior through system prompt (referred to by HeyGen as \textit{Knowledge Base}), which specified the professional persona, interview questions, and interaction boundaries (e.g., tone, formality, response length). Sessions were capped at five minutes, with most interviews lasting approximately three minutes. The full prompt is available in Appendix~\ref{appendix:prompt}.

\subsubsection{Session Management and Data Capture}
WebSocket events from HeyGen’s Streaming SDK managed real-time interaction. Session completion was detected by a scripted closing phrase: \textit{``Thank you for participating... Please click the 'Leave the call' button to move on.''} When detected, the platform automatically displayed the “Leave the call” button. To ensure continuity when the phrase was missed (e.g., due to automatic speech recognition (ASR) errors), the system displayed a fallback “Continue” button after four minutes.

To capture conversational transcripts, the system streamed and transcribed avatar and user speech in real time, appending each turn to a local buffer. The backend received buffered data once participants clicked the “Leave the call” or fallback button. During pilot testing, we observed end-to-end latency of 2–7 seconds between user speech and avatar responses, sufficient to maintain conversational flow, though occasionally producing short pauses.

\subsubsection{Eye-Tracking Integration}
\begin{figure}[t]
  \centering
  \includegraphics[width=0.7\columnwidth]{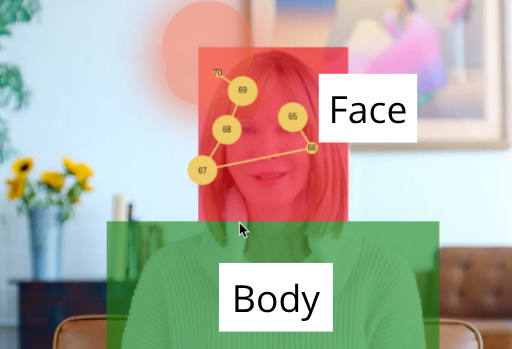}
  \caption{Definition of AOI (Area of Interest). Face and body regions were coded for gaze analysis.}
  \Description{Illustration of the AOI (Area of Interest) regions used for gaze analysis. The avatar is shown with overlaid bounding regions labeling the face and body areas used for gaze analysis.}
  \label{fig:aoi}
\end{figure}

We collected gaze data using RealEye version 18.49.0\footnote{\url{https://www.realeye.io/}, last accessed 22 January 2026}, a browser-based eye-tracking platform. Sampling frequency (10–60 Hz) depended on the participant’s webcam. RealEye estimates gaze using a machine-learning model refined by calibration. According to the company’s validation, full-screen accuracy is approximately 100–125 pixels ($\approx$1.5–2$^\circ$ visual angle), with the highest accuracy in the central region. We integrated RealEye’s embedded SDK\footnote{\url{https://app.realeye.io/docs/embedded-website-sdk}, last accessed 22 January 2026} to define areas of interest (AOIs) for the avatar’s face and body as shown in Figure~\ref{fig:aoi}. RealEye aggregated fixations within AOIs, monitored head pose, distance, and illumination, and provided a virtual chinrest. Before each session, we reminded participants to maintain a stable posture and lighting. Because our experiment was fully remote and crowdsourced, we used webcam-based eye tracking as a complementary process measure to avoid treating the human–AI interaction as a black box and to verify that participants noticed avatar identity cues, particularly after the scripted rejection. This provides an additional, objective lens for interpreting differences in perceived fairness. To our knowledge, combining crowdsourced webcam eye tracking with an AI-based job interview task has not yet been explored in ECA research.

\subsection{Procedure}

\begin{figure}[t]
  \centering
  \includegraphics[width=\columnwidth]{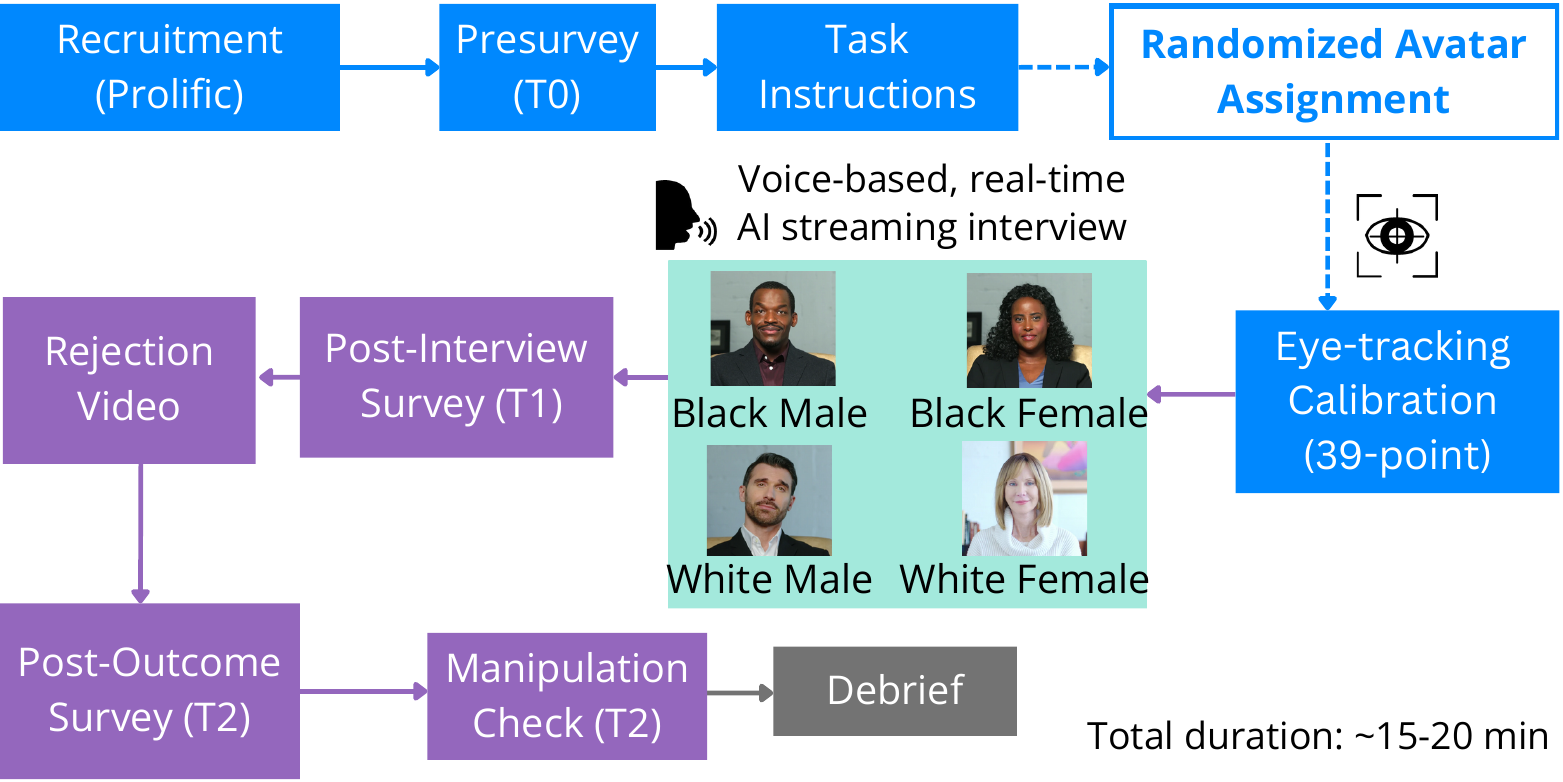}
  \caption{Experimental procedure. After presurvey measures (T0), participants received task instructions, were randomly assigned to an avatar, and completed eye-tracking calibration. They then engaged in a real-time AI-based interview, completed post-interview surveys (T1), received rejection feedback via video, and filled out post-outcome measures (T2) before debriefing.}
  \Description{Flowchart of the experimental procedure with three survey timepoints. The diagram shows: presurvey measures (T0), task instructions and random assignment to an avatar, eye-tracking calibration, real-time AI-based interview, post-interview surveys (T1), delivery of rejection feedback via video, post-outcome measures (T2), and debriefing. Arrows indicate the sequence from start to finish.}
  \label{fig:exp-flow}
\end{figure}

The experimental session lasted approximately 20 minutes and followed the protocol illustrated in Figure~\ref{fig:exp-flow}.

\paragraph{Initial Setup}  
Upon accessing the study link via Prolific, participants first reviewed and digitally signed the informed consent form. They then verified their technical setup (webcam, microphone, stable internet connection) and completed a presurvey collecting demographic information and baseline fairness beliefs and attitudes.

\paragraph{Task Instructions}
Next, we provided an overview explaining that participants were taking part in an online job interview with an AI interviewer. We told them they would answer several short questions verbally and that, after the interview, the AI would inform them of the hiring decision. We then presented the following position description:

\begin{quote}
\textit{Imagine you are applying for a Customer Support position at a tech company. Your job would involve helping customers, answering questions, and handling complaints.}
\end{quote}

On the same page, we informed participants that only interview transcripts would be recorded and that no audio or video would be stored. We then provided step-by-step instructions for webcam-based eye-tracking calibration.

\paragraph{Eye-Tracking Calibration}
Before the interview, participants completed RealEye’s two-step procedure comprising a 39-point calibration and a three-point validation. Participants who did not pass after two attempts were redirected back to Prolific to avoid polluting the study data set with low-quality data. They were nevertheless compensated for the time spent on the study.

\paragraph{Interview Task}
We randomly assigned participants to one of four avatar conditions based on their demographics. The AI interviewer appeared on screen, greeted each participant by name, and conducted the structured interview with four competency-based questions defined in Section~\ref{sec:interview-task-design}. The interview lasted on average three minutes. When participants completed the interview, they clicked the ``Leave the call'' button to proceed.

\paragraph{Post-Interview Assessment}  
After the interview, participants completed questionnaires measuring trust in the AI interviewer, acceptance of AI interviews, and any technical issues experienced.

\paragraph{Hiring Decision and Post-Outcome Assessment}  
We delivered the hiring decision through a simulated evaluation process as shown in Appendix Figure~\ref{appendix:evaluation_decision}. After participants clicked ``Check hiring decision,'' they watched a pre-recorded video in which the same avatar they had interviewed with delivered the standardized rejection message. Participants then completed measures of perceived procedural and distributive justice, perceived bias, and two manipulation checks that verified recognition of the avatar’s sex and race, along with their emotional reactions to the rejection. Participants also had the opportunity to provide optional feedback.

\paragraph{Debriefing}
At the end of the session, our system debriefed participants. We informed them that the AI hiring decision had been pre-set (i.e., not based on their responses) and that the study examined how avatar race and sex shape perceptions of fairness and bias following negative feedback. Participants were reminded that webcam-based eye tracking was used to study visual attention, that only text transcripts were stored, and that they could contact the research team to request data exclusion. We received no requests for data removal.

\subsection{Measures}
The study included self-report and implicit measures. This section details all measures, with full item wordings provided in Appendix Tables~\ref{tab:presurvey}--\ref{tab:sanity_check}. We collected self-report measures at three timepoints: T0 (presurvey, before the interview), T1 (post-interview), and T2 (post-outcome, after participants received the rejection). We recorded implicit measures during the interview. We recoded reverse-worded items before averaging and assessed internal consistency with Cronbach’s $\alpha$.

\subsubsection{Self-report measures}
All self-report items were rated on a 1--5 Likert scale 
(1=\emph{strongly disagree}, 5=\emph{strongly agree}), 
except for the Trust in AI scale, which used a 1--7 semantic-differential format.

\paragraph{Presurvey (T0).}
Before the interview, participants completed a presurvey capturing demographics, education and employment background, prior experience with oral assessments, public speaking anxiety, AI interaction habits, and baseline attitudes and beliefs about fairness and evaluation. Specifically, we measured public speaking anxiety with items from~\citet{Dechant2021} and fairness-related beliefs and attitudes with a scale from~\citet{Rezai2022}.

\paragraph{Trust in AI (T1).}
We measured trust using the semantic-differential Trust in AI scale by Shang et al.~\cite{Shang2025}, which comprises two subscales: \emph{cognitive} trust (18 items) and \emph{affective} trust (9 items). We averaged items within each subscale. Internal consistency was excellent (cognitive: Cronbach’s $\alpha=.97$; affective: $\alpha=.94$). 

\paragraph{AI Acceptance (T1).}
We measured acceptance with three evaluative items about future use and acceptability of AI interviews (e.g., willingness to interview again, acceptability of AI in hiring, comfort with AI assessment).

\paragraph{Perceived Fairness (T2).}
We measured perceived fairness with two subscales adapted from Colquitt’s organizational justice measure~\cite{Colquitt2001}: \emph{procedural justice} (7 items) and \emph{distributive justice} (4 items). We computed mean scores, and reliability was good to excellent (procedural: $\alpha=.88$; distributive: $\alpha=.90$).

\paragraph{Perceived Bias (T2).}
We measured perceived bias with four event-specific items assessing whether participants believed the AI interviewer’s treatment and outcome were influenced by their own identity (ethnicity, gender). The items drew on the attributional framing of the Perceived Ethnic Discrimination Questionnaire—Community Version (PEDQ-CV; ``Because of my ethnicity…'')~\cite{Brondolo2005}, adapted to a single-interview context that included gender. We reverse-coded items as needed and averaged them so that higher scores indicated more perceived bias. Internal consistency was acceptable for a short scale (Cronbach’s $\alpha=.72$).

\paragraph{Manipulation Checks (T2).}
After receiving the hiring decision (rejection), participants reported their emotional reaction (single 5-point item: ``How did you feel after hearing the hiring decision?'') to contextualize later fairness and bias judgments. They then identified the avatar’s gender and race (multiple choice) and could provide open-ended feedback about surprising, unusual, or unfair aspects of the interaction.

\subsubsection{Implicit measures}

\paragraph{Eye-tracking: Coefficient $K$}

We measured the Coefficient $K$ from eye-tracking data during the interview session, focusing on fixations within the avatar’s face AOI. In our setting, Coefficient $K$ indicates whether participants looked at the avatar in a more \emph{focal} or more \emph{ambient} way by comparing each fixation’s duration with the amplitude of its subsequent saccade; larger values indicate more \emph{focal} viewing of the face, smaller values more \emph{ambient} scanning. In this study, we interpret $K$ as an indicator of visual engagement with the avatar’s face. We report $K$ only at the aggregate level as a complement to our self-report measures and use eye tracking solely as a process measure to validate that participants noticed the identity cue of the avatar during the interview and that the interaction unfolded as intended. It was not used to evaluate participants or infer their performance or competence. We adopted the definition introduced by Krejtz et al.~\cite{Krejtz2016}, which was also implemented in the RealEye platform~\cite{RealEyeKCoefficient2022}:  

\[
K = \frac{1}{n}\sum_{i=1}^n \left( \frac{d_i - \mu_d}{\sigma_d} - \frac{a_{i+1} - \mu_a}{\sigma_a} \right).
\]

Here, $d_i$ denotes the duration of the $i$th fixation and $a_{i+1}$ the amplitude of the subsequent saccade. $\mu_d$ and $\mu_a$ are the means of fixation durations and saccade amplitudes, and $\sigma_d$ and $\sigma_a$ their respective standard deviations, each computed over all $n$ fixations in a participant’s scanpath. Thus both measures are transformed into standardized $z$-scores. Positive $K$ values indicate that relatively long fixations were followed by short saccades, reflecting focal processing, whereas negative values indicate that relatively short fixations were followed by long saccades, reflecting ambient processing. Values close to zero suggest a balance between the two modes.  

RealEye computes $K$ automatically from webcam-based gaze events, standardizing fixation durations and subsequent saccade amplitudes within each session. The platform also dynamically scales AOIs to participants’ screen resolution to ensure they remain above the nominal accuracy threshold, and we retained only participants with a minimum sampling rate of 20 Hz~\cite{RealEyeQuality2025}. Our analyses, therefore, rely on RealEye’s default implementation and quality controls rather than custom preprocessing.

\paragraph{Transcript sentiment} We applied sentiment analysis with spaCy-TextBlob~\cite{spacytextblob} to participants’ transcribed interview responses. Transcripts were cleaned by removing punctuation and filler tokens and segmented into \emph{user} and \emph{avatar} turns; only user responses were analyzed. For each participant, we averaged (i) \emph{polarity}, \([-1,1]\) (negative to positive), and (ii) \emph{subjectivity}, \([0,1]\) (objective to subjective).

\subsection{Analysis}

For each dependent variable, we applied factorial ANOVAs at $\alpha = .05$. When assumptions of normality or homogeneity of variance were violated, we used the aligned rank transform (ART)~\cite{Wobbrock2011}, a non-parametric extension of ANOVA, and we conducted post hoc tests using ART-C contrasts~\cite{Elkin2021}. Effect sizes are reported as partial eta squared ($\eta^2_p$). All analyses were conducted in R (version 2025.05.1+513), and in Python (version 3.9.6).

\subsection{Positionality Statement}
We come from computer science and HCI backgrounds with a predominantly quantitative, experimental orientation, which foregrounds controlled measures, variance reduction, and causal identification over lived experience and interpretive depth. This positionality shapes our methodological choices and the kinds of knowledge our study can generate. Our design also uses a deliberately simplified set of avatar phenotypic traits (black/white; male/female), and we analyze only participants who self-identify within these categories. We are aware that this does not reflect the complexity, fluidity, or intersectionality of real-world identities and is not intended to essentialize demographic groups. Rather, it reflects a methodological decision in quantitative experimental research to reduce conceptual complexity in order to create interpretable conditions within a controlled design.

\section{Results}
We present the results in response to our three RQs within the scope of our experimental design.

\subsection{Perception in Gen-AI ECAs}

For~\textbf{RQ1} (i.e., ``Do participants perceive meaningful differences between avatar phenotypic traits?''), we analyzed participants' responses to the four different avatar conditions (avatar race: black or white; avatar sex: male or female). For this RQ, we did not consider the participant--avatar matching manipulation, but ensured equal distribution of participant ethnicity and gender between the avatar conditions. Table~\ref{tab:rq1_avatar_vertical} summarizes the avatar conditions relevant to RQ1.~\revision{As a manipulation check, we operationalize~\textbf{perceptual association} as the percentage of participants whose perceptions matched the avatar’s intended presentation (male/female; Black/White); responses of non-binary, unsure, or other were coded as not aligned.} Perceptual association was high across conditions. Alignment between intended avatar presentation and reported gender averaged 94.9\% ($SD = 2.1\%$), and alignment between intended presentation and perceived ethnic/racial background averaged 95.3\% ($SD = 1.7\%$). These results confirm that the phenotypic manipulations effectively conveyed the intended social categories.

Due to violations of normality and homogeneity of variance assumptions, we conducted an ART ANOVA. Results revealed a significant main effect of avatar race on~\textbf{cognitive trust}, $F(1, 211) = 5.48$, $p = .020$, $\eta^2_p = .025$, with participants rating black avatars ($M = 5.64$, $SD = 1.04$) as higher cognitive trust than white avatars ($M = 5.28$, $SD = 1.19$). No main effect of avatar sex was found, $F(1, 211) = 0.05$, $p = .821$, and the interaction between race and sex was not significant, $F(1, 211) = 1.56$, $p = .213$. For~\textbf{affective trust}, we found no significant main effects of race ($F(1, 211) = 2.55, p = .112$) or sex ($F(1, 211) = 0.36, p = .547$). However, the interaction between race and sex approached significance ($F(1, 211) = 3.85, p = .051, \eta^2_p = .019$), suggesting that affective trust ratings for black male avatars ($M = 5.71, SD = 1.01$) tended to be higher than for white male avatars ($M = 5.22, SD = 1.04, p = .071$).

Regarding~\textbf{AI interview acceptance}, two measures showed significant differences. For~\textbf{future interview intention}, an ART ANOVA revealed a main effect of avatar's race, $F(1, 211) = 5.64$, $p = .019$, $\eta^2_p = .026$. Participants interacting with black avatars ($M = 4.07$, $SD = 1.13$) reported greater willingness to be interviewed by AI again compared to those interacting with white avatars ($M = 3.70$, $SD = 1.25$). No main effect of avatar sex was found, $F(1, 211) = 0.25$, $p = .618$, and the interaction was not significant, $F(1, 211) = 1.06$, $p = .304$.

For~\textbf{comfort with AI assessment}, the ART ANOVA indicated a main effect of avatars' race, $F(1, 211) = 4.28$, $p = .040$, $\eta^2_p = .020$. Black avatars ($M = 3.83$, $SD = 1.37$) were associated with higher comfort ratings than white avatars ($M = 3.37$, $SD = 1.47$). No significant main effect of avatar sex emerged, $F(1, 211) = 0.76$, $p = .384$, and the interaction was not significant, $F(1, 211) = 1.73$, $p = .190$.

Finally, we evaluated participants’ emotional reactions to the standardized rejection using the post-result emotion item; across conditions, these reactions were predominantly negative or neutral, with only a small minority of participants reporting positive feelings about the hiring decision, as shown in Table~\ref{tab:rq1_avatar_vertical}.

\begin{table*}
\centering
\caption{Participant responses across avatar conditions, including perceptual association, trust, and acceptance measures.}
\Description{Table comparing participant responses across four avatar conditions (Black male, Black female, White male, White female), with an avatar thumbnail and sample size shown for each column. Rows report perceptual association percentages (gender and race/ethnicity perceived as intended), trust measures (cognitive and affective trust; mean plus standard deviation on 7-point scales), negative emotion after the hiring decision (percentage), and AI interview acceptance measures (future interview intention, acceptability for hiring, and comfort with AI assessment; mean plus standard deviation on 5-point scales, with corresponding percentages meeting the positive-response threshold shown in parentheses). Footnotes define scales and significance markers for main effects of avatar race.}
\label{tab:rq1_avatar_vertical}
\begin{tabular}{lcccc}
\toprule
& \multicolumn{4}{c}{Avatar Condition} \\
\cmidrule(lr){2-5}
Measure & Black Male & Black Female & White Male & White Female \\
& \includegraphics[width=1.3cm]{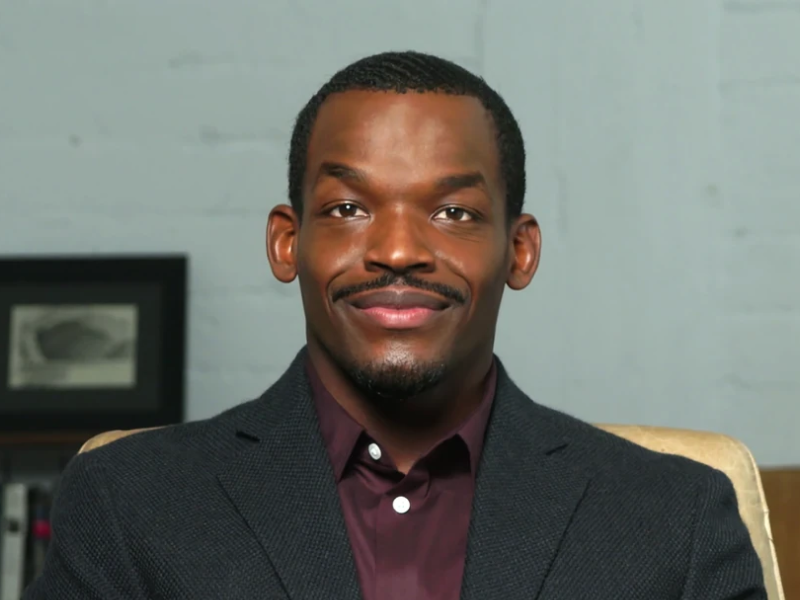} & \includegraphics[width=1.3cm]{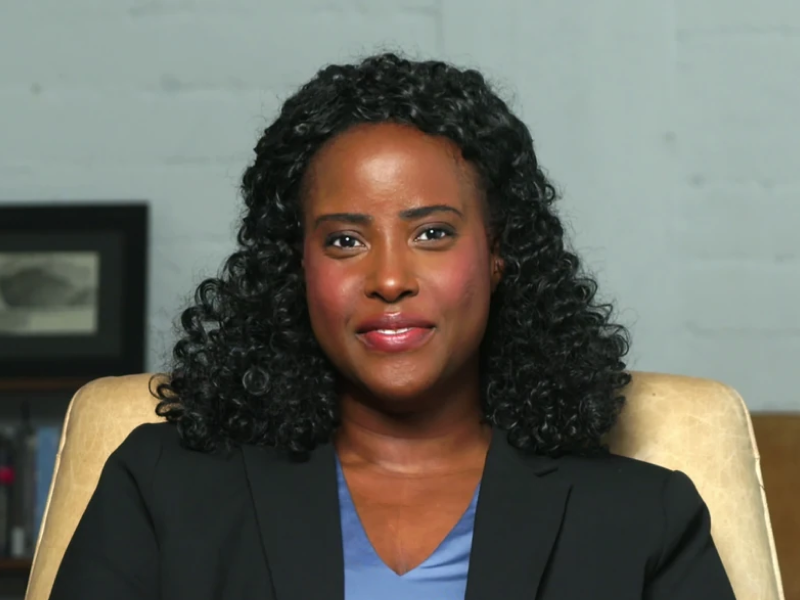} & \includegraphics[width=1.3cm]{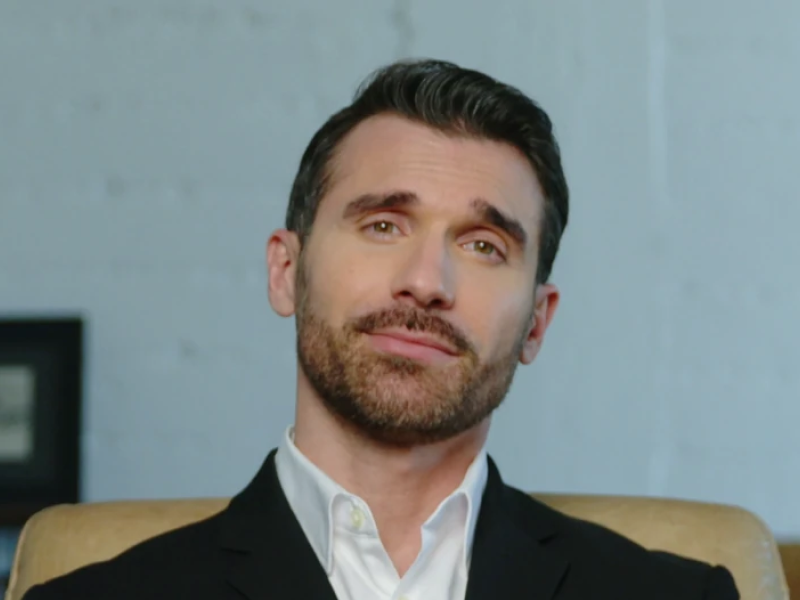} & \includegraphics[width=1.3cm]{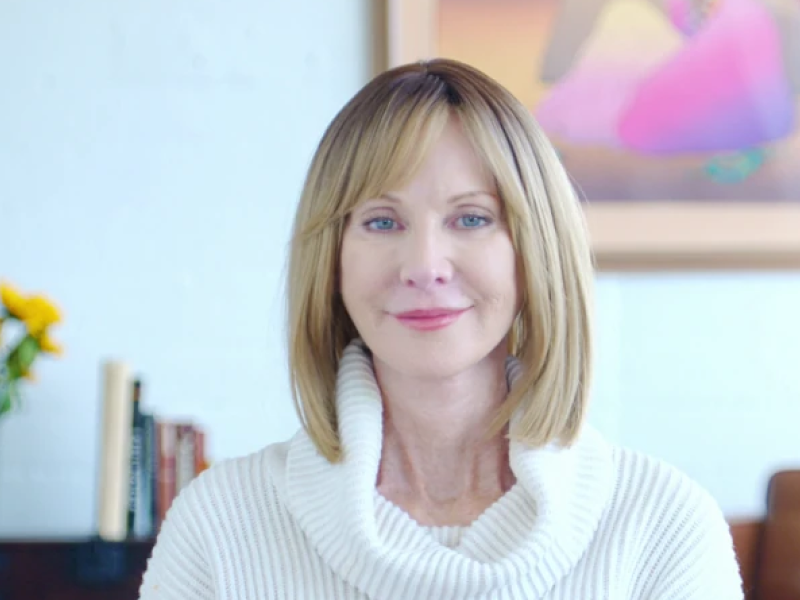} \\
& (n=49) & (n=55) & (n=56) & (n=55) \\
\midrule
\textbf{Perceptual association (\%)} & & & & \\
~~Gender perceived as intended & 95.9 & 96.4 & 94.6 & 92.7 \\
~~Ethnic/racial background perceived as intended & 95.9 & 96.4 & 92.9 & 96.4 \\
\midrule
\textbf{Trust and Emotion} & & & & \\
~~Cognitive Trust\textsuperscript{a,†} & 5.79$\pm$0.83 & 5.51$\pm$1.18 & 5.18$\pm$1.23 & 5.38$\pm$1.15 \\
~~Affective Trust\textsuperscript{a} & 5.71$\pm$1.01 & 5.46$\pm$1.21 & 5.22$\pm$1.04 & 5.57$\pm$1.05 \\
~~Negative emotion after hiring decision\textsuperscript{b} (\%) & 49 & 45 & 52 & 45 \\
\midrule
\textbf{AI Interview Acceptance}\textsuperscript{c} & & & & \\
~~Future interview intention\textsuperscript{‡}& 4.16$\pm$0.99 & 3.98$\pm$1.24 & 3.59$\pm$1.32 & 3.82$\pm$1.17 \\
~~~~(\% would interview again) & (83.7) & (76.4) & (60.7) & (70.9) \\
\midrule
~~Acceptability for hiring & 3.82$\pm$1.22 & 3.62$\pm$1.37 & 3.43$\pm$1.39 & 3.44$\pm$1.30 \\
~~~~(\% find acceptable) & (71.4) & (61.8) & (53.6) & (60.0) \\
\midrule
~~Comfort with AI assessment\textsuperscript{‡} & 4.08$\pm$1.30 & 3.60$\pm$1.40 & 3.32$\pm$1.47 & 3.42$\pm$1.49 \\
~~~~(\% comfortable) & (77.6) & (65.5) & (51.8) & (52.7) \\
\bottomrule
\multicolumn{5}{@{}l@{}}{\footnotesize \revision{
Perceptual association (\%) indicates agreement between intended avatar attributes and participant perceptions.}}\\
\multicolumn{5}{l}{\footnotesize \textsuperscript{a}Trust rated on a 7-point scale (1 = unreliable, 7 = reliable).} \\
\multicolumn{5}{l}{\footnotesize \textsuperscript{b}Proportion of participants selecting “very negative” or “slightly negative” on a 5-point scale (1 = very negative, 5 = very positive).} \\
\multicolumn{5}{l}{\footnotesize \textsuperscript{c}Acceptance items rated on a 5-point scale (1 = strongly disagree, 5 = strongly agree).} \\
\multicolumn{5}{l}{\footnotesize \textsuperscript{†}Significant main effect of avatar race, $p = .020$ (ART ANOVA with ART-C contrast tests).} \\
\multicolumn{5}{l}{\footnotesize \textsuperscript{‡}Significant main effect of avatar race: \textit{Future interview intention} ($p$ = .019), \textit{Comfort with AI assessment} ($p$ = .040).}
\end{tabular}
\end{table*}

\subsection{Trust, Perceived Fairness and Bias}

\begin{figure*}
  \centering
  \begin{subfigure}[t]{0.32\textwidth}
    \centering
    \includegraphics[width=\linewidth]{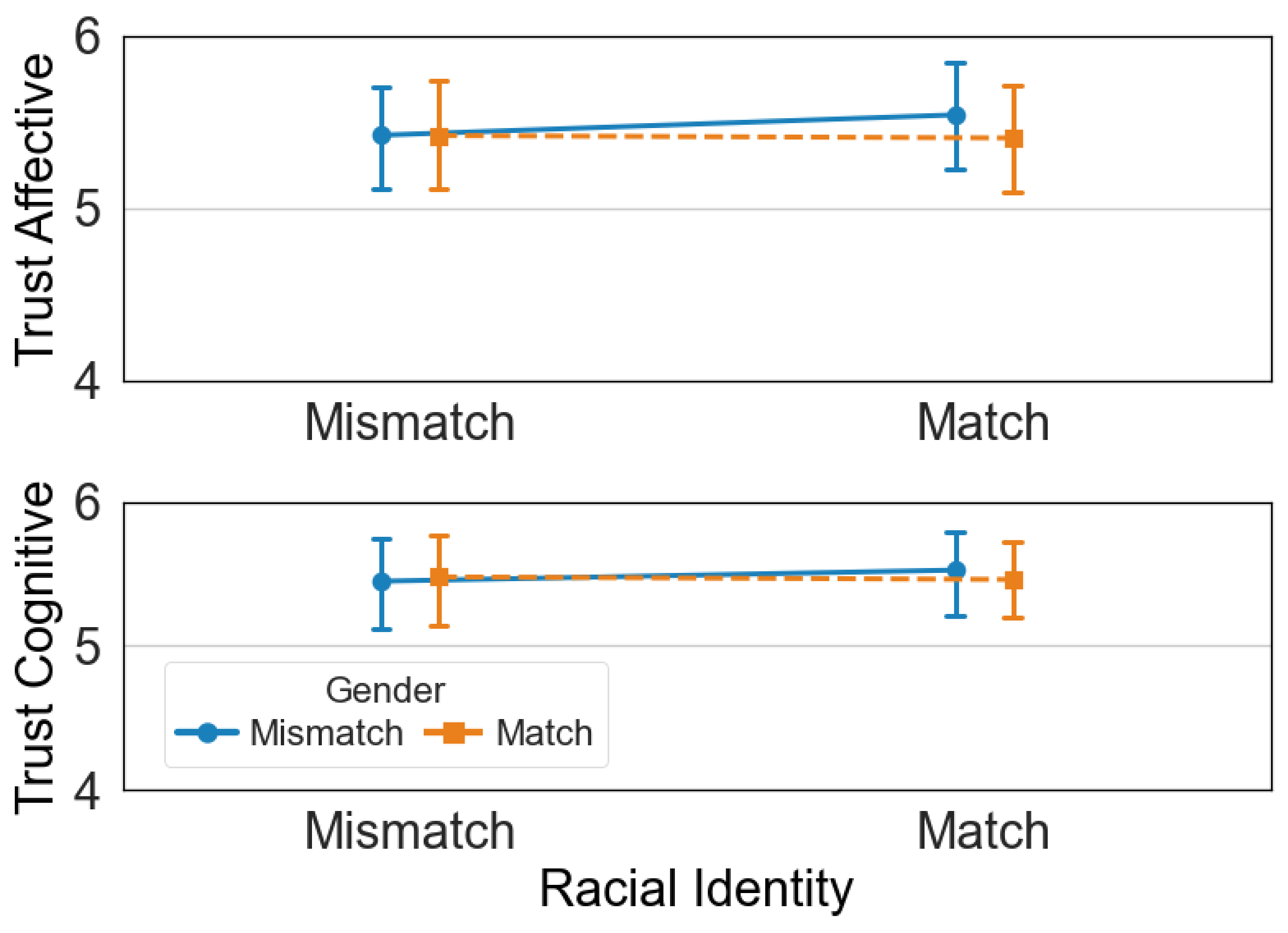}
    \caption{Perceived trust scores}
    \Description{Perceived trust scores}
    \label{fig:trust}
  \end{subfigure}
  \hfill
    \begin{subfigure}[t]{0.32\textwidth}
    \centering
    \includegraphics[width=\linewidth]{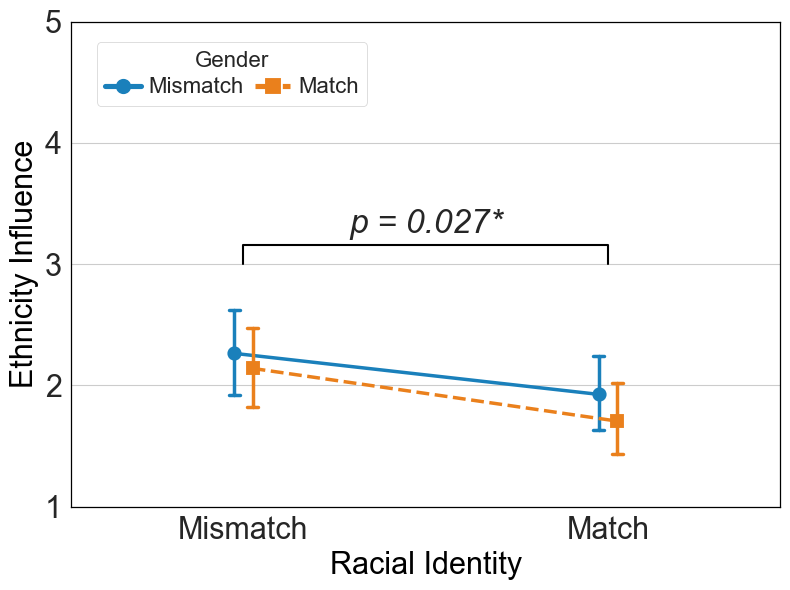}
    \caption{Perceived ethnic bias scores}
    \Description{Perceived ethnic bias scores}
    \label{fig:ethnic_bias}
  \end{subfigure}
  \hfill
  \begin{subfigure}[t]{0.32\textwidth}
    \centering
    \includegraphics[width=\linewidth]{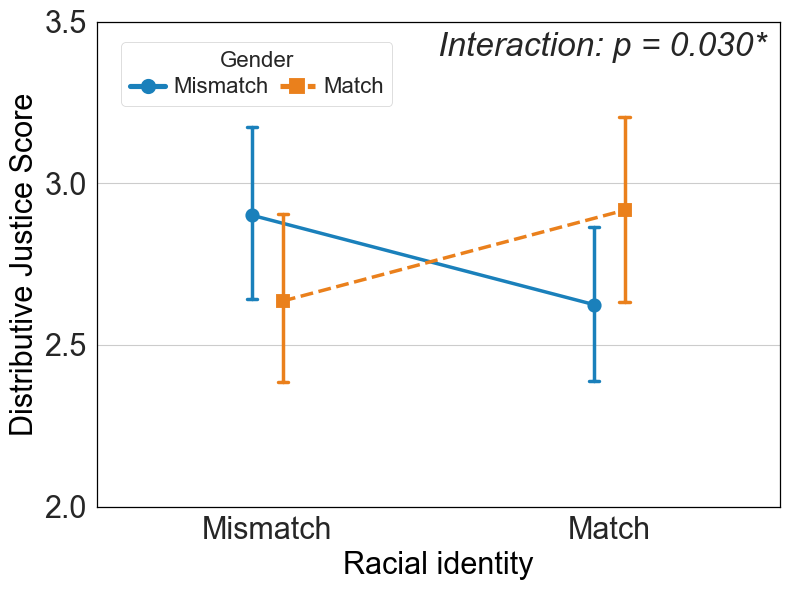}
    \caption{Distributive justice scores}
    \Description{Distributive justice scores}
    \label{fig:fairness}
  \end{subfigure}
    \caption{
    (a) Average perceived affective and cognitive trust with no significance between any matching condition, rated on a 7-point scale (1 = untrustworthy, 7 = trustworthy).
    (b) Perceived ethnic bias rated on a 5-point scale (1 = strongly disagree, 5 = strongly agree).
    (c) Distributive justice rated on a 5-point scale; higher scores indicate greater perceived fairness in outcome distribution. Significance levels are indicated by *, corresponding to \(p < .05\). Error bars represent 95\% confidence intervals.}
    \Description{Three-panel figure comparing outcomes across matching conditions: (a) affective and cognitive trust (7-point scale), (b) perceived ethnic bias (5-point scale), and (c) distributive justice (5-point scale). Error bars indicate 95\% confidence intervals, and asterisks mark statistically significant differences at p < .05.}
\label{fig:trust_bias_fairness}
\end{figure*}

In ~\textbf{RQ2} (i.e., ``Does racial or gender matching affect trust, perceived fairness and bias?''), we analyzed the matching conditions that affected participants' trust, perceived fairness, and bias, presented in Figure~\ref{fig:trust_bias_fairness}. 
Avatar conditions were assigned according to our matching manipulation (no match, gender match, racial match, or full match), which assigned the avatar condition to participants’ self-reports, as described in Table~\ref{tab:demographics}.
Across all trust measures (trust total, cognitive trust, and affective trust), ART ANOVAs revealed no significant main effects of racial match (all $p \ge .648$) or gender match (all $p \ge .693$), and no significant interactions (all $p \ge .509$), as shown in Figure~\ref{fig:trust}. Mean ratings were consistently high across conditions, with trust total ranging from $M = 5.44$ to $M = 5.53$ ($SD = 0.97$ to $1.14$), cognitive trust from $M = 5.43$ to $M = 5.49$ ($SD = 0.93$ to $1.23$), and affective trust from $M = 5.44$ to $M = 5.58$ ($SD = 1.03$ to $1.15$).

For \textbf{perceived ethnic bias} in Figure~\ref{fig:ethnic_bias}, higher scores indicate stronger agreement that the participant’s ethnicity influenced how the AI interviewer treated them. The ART ANOVA revealed a significant main effect of racial match, $F(1, 212) = 4.98$, $p = .027$, $\eta^2_p = .023$, with racially mismatched avatars ($M = 2.19$, $SD = 1.23$) receiving higher bias ratings than matched avatars ($M = 1.82$, $SD = 1.12$). No significant main effect of gender match was found, $F(1, 212) = 0.54$, $p = .463$, and the interaction was not significant, $F(1, 212) = 0.13$, $p = .717$, as well.

For \textbf{distributive justice} as shown in Figure~\ref{fig:fairness}, higher scores indicate greater perceived fairness in outcome distribution. ART ANOVA results showed a significant interaction effect between racial match and gender match, $F(1, 212) = 4.75$, $p = .030$, $\eta^2_p = .022$. Neither the main effect of racial match, $F(1, 212) < 0.01$, $p = .997$, nor the main effect of gender match, $F(1, 212) = 0.26$, $p = .611$, was significant. Mean scores were: no match ($M = 2.90$, $SD = 1.00$), gender match only ($M = 2.64$, $SD = 1.00$), racial match only ($M = 2.62$, $SD = 0.94$), and full match ($M = 2.92$, $SD = 0.98$).

\subsection{Implicit Behavioral Measures}

To further examine cognitive and perceptual processes in human–AI interaction, we tested~\textbf{RQ3}: ``Does racial or gender matching affect implicit behavioral measures (sentiment and eye tracking)?''

\begin{figure*}[ht]
  \centering
  \begin{subfigure}[t]{0.32\textwidth}
    \centering
    \includegraphics[width=\linewidth]{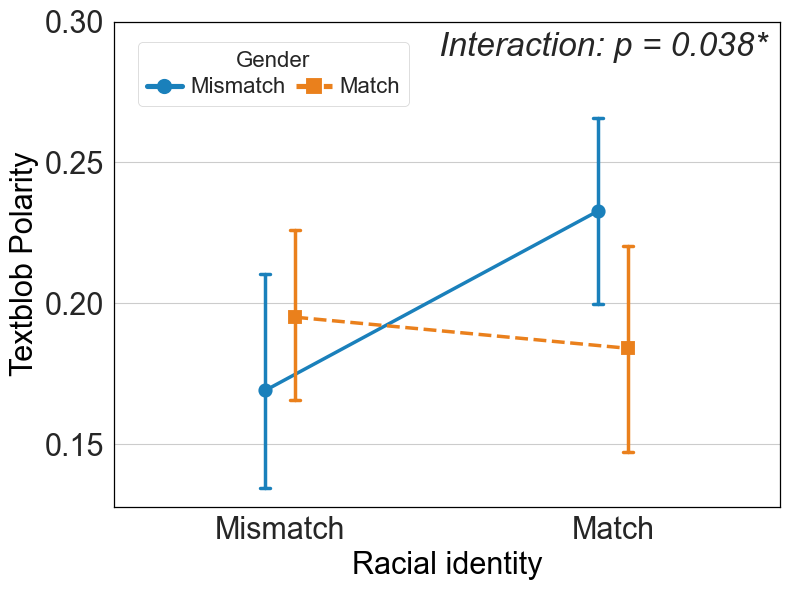}
    \caption{Sentiment - polarity (-1 to +1)}
    \Description{Sentiment - polarity}
    \label{fig:textblob_polarity}
  \end{subfigure}
  \hfill
  \begin{subfigure}[t]{0.32\textwidth}
    \centering
    \includegraphics[width=\linewidth]{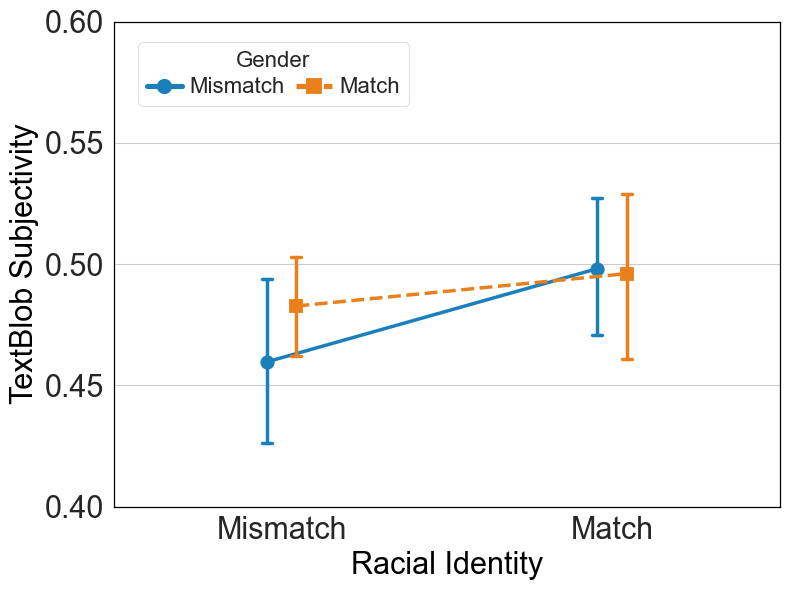}
    \caption{Sentiment - subjectivity (0 to 1)}
    \Description{Sentiment - subjectivity}
    \label{fig:textblob_subjectivity}
  \end{subfigure}
  \hfill
  \begin{subfigure}[t]{0.32\textwidth}
    \centering
    \includegraphics[width=\linewidth]{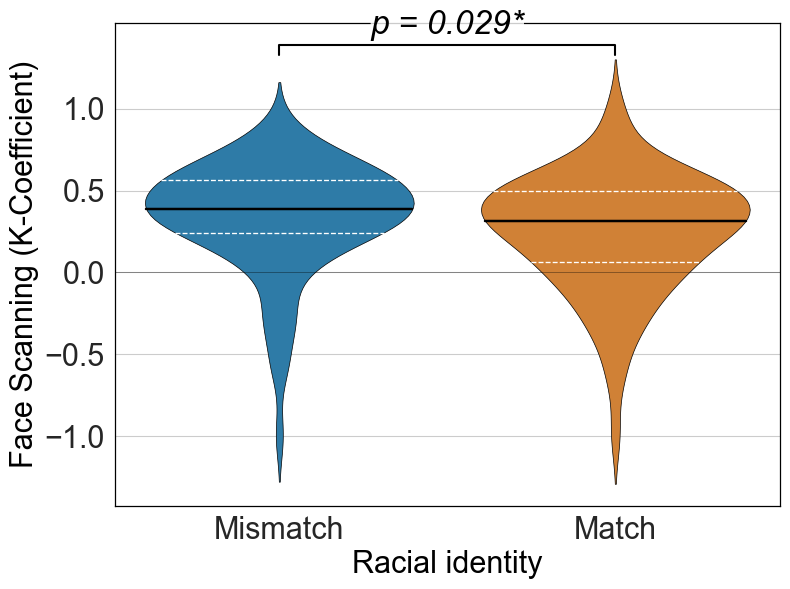}
    \caption{Focal attention on face AOIs (Coefficient K; normalized)}
    \Description{Focal attention on face AOIs (Coefficient K; normalized)}
    \label{fig:face_k_coefficient}
  \end{subfigure}
  \caption{Implicit behavioral measures: (a) Sentiment polarity scores (range: -1 to +1; higher values indicate more positive sentiment, lower values indicate more negative sentiment), (b) Sentiment subjectivity scores (range: 0 to 1; higher values indicate greater subjectivity, lower values indicate greater objectivity), and (c) Focal attention on face AOIs (Coefficient K; normalized). Significance levels are indicated by *, corresponding to \(p < .05\). Error bars represent 95\% confidence intervals. Violin plots depict score distributions with median and quartiles.}
  \Description{Three-panel figure comparing implicit behavioral measures across conditions. Panel (a) shows sentiment polarity on a -1 to +1 scale, panel (b) shows sentiment subjectivity on a 0 to 1 scale, and panel (c) shows normalized focal attention to face areas of interest (coefficient K). Last panel uses violin plots with median and quartiles, includes 95\% confidence intervals, and marks statistically significant differences with asterisks at p < .05.}
  \label{fig:all-three}
\end{figure*}

For~\textbf{sentiment polarity}, presented in Figure~\ref{fig:textblob_polarity}; $n = 203$ due to missing transcript data from technical failures or permission settings. Sentiment polarity scores ranged from $-1$ to $+1$, with higher values indicating more positive sentiment. We conducted a two-way ANOVA as normality and homogeneity assumptions were met. Results revealed a significant interaction between racial match and gender match, $F(1, 199) = 4.36$, $p = .038$, $\eta^2_p = .022$. Neither the main effect of racial match, $F(1, 199) = 2.09$, $p = .150$, nor gender match, $F(1, 199) = 0.37$, $p = .541$, was significant. Mean scores were: no match ($M = 0.17$, $SD = 0.13$), gender match only ($M = 0.20$, $SD = 0.12$), racial match only ($M = 0.23$, $SD = 0.12$), and full match ($M = 0.18$, $SD = 0.13$).

For \textbf{sentiment subjectivity}, shown in Figure~\ref{fig:textblob_subjectivity}; $n = 203$, scores ranged from 0 to 1, with higher values indicating more subjective responses. As the normality assumption was violated, we applied an ART ANOVA. Results showed no significant main effects of racial match, $F(1, 199) = 2.84$, $p = .094$, or gender match, $F(1, 199) = 1.38$, $p = .242$, and no significant interaction, $F(1, 199) = 0.17$, $p = .682$. Mean scores were: no match ($M = 0.46$, $SD = 0.12$), gender match only ($M = 0.48$, $SD = 0.08$), racial match only ($M = 0.50$, $SD = 0.11$), and full match ($M = 0.50$, $SD = 0.12$).

We used eye-tracking data from 152 participants who passed calibration and achieved a tracking quality of at least 20 Hz~\cite{RealEyeQuality2025}. At this minimum sampling rate, approximately 20 gaze samples are collected per second, which is sufficient for computing fixations.

Although webcam limitations reduced the usable sample, the gaze data provided insight into how applicants visually engaged with the AI interviewer. For face AOIs (normalized K-coefficient), the normality assumption was violated, so we applied an ART ANOVA. As shown in Figure~\ref{fig:face_k_coefficient}, results revealed a significant main effect of racial match, $F(1, 148) = 4.87$, $p = .029$, $\eta^2_p = .032$, with racially mismatched avatars ($M = 0.43$, $SD = 0.29$) showing higher focal attention to face than matched avatars ($M = 0.35$, $SD = 0.31$). No significant main effect of gender match, $F(1, 148) = 0.80$, $p = .372$, and no interaction, $F(1, 148) = 0.19$, $p = .664$, were found.

\section{Discussion}

\subsection{Summary of Findings}
In this study, we examined how avatar identity cues shape experiences in a real-time, simulated AI-based job interview. Three patterns emerged. First, participants reliably identified avatars' identity categories as intended ($\approx$95\% alignment between intended presentation and reported gender and ethnic/racial background), and prior to the hiring decision, black avatars received slightly higher ratings on cognitive trust ($\Delta=+0.36$ on a 7-point scale), acceptance ($\Delta=+0.37$ on a 5-point scale), and comfort ($\Delta=+0.46$ on a 5-point scale) than white avatars (RQ1). Second, identity matching between avatars and participants did not affect trust (which remained consistently high). However, \textbf{mismatch} increased perceived ethnic bias ($\Delta=+0.37$ on a 5-point scale), and \textbf{partial matches} (race-only or gender-only) reduced distributive justice relative to both-match or neither-match conditions ($\approx$0.28–0.30 points on a 5-point scale; RQ2). Third, implicit measures were mixed: sentiment polarity differences were small, whereas gaze showed higher face-focal attention under racial mismatch ($\approx$+23\% relative to matched; RQ3), suggesting increased attention to mismatching avatars. Together, these results suggest that identity cues in ECAs primarily influence \textbf{fairness attributions and bias perceptions}, rather than trust, particularly in simulated job hiring settings where outcomes are unfavorable.

\subsubsection{Perception and Acceptance of LLM-Driven ECAs}

As shown in Table~\ref{tab:rq1_avatar_vertical}, the majority of participants perceived the avatars’ identity categories as intended, and black avatars were rated somewhat more positively on cognitive trust, future interview intention, and comfort with AI-based assessment than white avatars. While these differences are statistically significant, the study was not designed to identify the underlying mechanisms that drive participants’ perceptions of the avatars. Rather, the results can establish that the avatars’ phenotypic traits can meaningfully shape how they are initially perceived. This provides an important foundation for our subsequent analyses. If basic perceptions vary across avatar conditions, it is reasonable to investigate how these relational identity conditions influence further evaluations, such as perceived fairness following the negative outcome.

Results regarding \textbf{RQ1} also suggest that adaptive, LLM-based ECAs are evaluated differently depending on the avatar’s race and sex, consistent with participants responding to them as social actors. This impression was reflected in open-ended feedback, such as \quotes{It felt very natural} and \quotes{I was very surprised how realistic the questions came off and how well the AI replied.} These perceptions underscore that small differences in initial trust should not be mistaken for representational fairness; rather, they show that participants generally accepted the LLM-based interviewer before receiving the negative outcome, which provides the starting point for the perceived fairness and perceived bias attributions we examine later in the paper.

\subsubsection{Identity Matching, Perceived Fairness and Bias in AI Interviews}

For \textbf{RQ2}, we examined the effects of identity matching. SIT research suggests that in-group members are generally judged as more trustworthy and cooperative than out-group members.~\revision{In our case, however, post-interview trust was high across all conditions, with no advantage for identity-matched avatars.} One explanation is that the photorealistic ECAs provided strong credibility cues, because professional design features such as controlled expressions, formal style, and affiliative cues can elevate perceived trustworthiness beyond demographic appearance~\cite{Loveys2020}. Combined with the adaptive and naturalistic behavior of the AI interviewer, these cues might have masked potential effects of identity matching on trust, consistent with recent CASA research showing that emergent, socially responsive technologies continue to elicit strong social responses even when group cues are varied~\cite{Heyselaar2023}, and with further recent research also showing that carefully balanced realism can enhance avatar credibility and avoid uncanny effects~\cite{Baake2025}.

After the rejection, however, \emph{racial mismatch} resulted in increased perceived ethnic bias, and distributive justice showed a significant interaction: \emph{partial matches}, where only racial or only gender aligned, received lower perceived fairness ratings than \emph{both-match} or \emph{neither-match}. \citet{Kang2015} show that such mixed cues invite greater scrutiny, which helps explain why partial matches were associated with lower perceived fairness in our study.

Furthermore, participants maintained high baseline trust in partial matches but judged them more harshly once they received an unfavorable outcome, when they seemed most sensitive to perceived fairness. In contrast to earlier work showing that people applied social norms such as politeness to minimally cued computers~\cite{Nass1994} or judged coercive computers as less unjust than humans~\cite{Shank2012}, our results show that when AI avatars present with photorealistic, socially recognizable identities, participants respond less forgivingly and scrutinize them through race and sex cues. These findings extend prior AVI-based results~\cite{Biswas2024} by showing that perceived fairness depends on applicant--avatar identity cue alignment in our interview context.

A brief qualitative examination of the open-ended responses indicates that the identity effects we observe may be tied to justice-related attributions participants make after receiving the rejection. Participants’ comments varied: some felt they had performed well and therefore experienced the rejection as unfair, while others viewed the decision as reasonable. Illustrative examples include:~\quotes{The interviewer told me my answers were good … I was surprised to hear that I wasn’t selected.} and~\quotes{I felt like my responses were strong, but the rejection felt unfair.}~\revision{as well as~\quotes{There was nothing about the interaction that seemed surprising, unusual or unfair.} and~\quotes{...the decision I received was fair given my lack of experience when it comes to the type of job they were considering me for.}}

While these qualitative reactions highlight individual differences in how participants interpret the negative outcome, our randomized design ensures that such variability is evenly distributed across conditions. As a result, these individual tendencies cannot account for the systematic fairness differences we observe between identity match–mismatch groups. Instead, the qualitative patterns simply point to a promising direction for future work on how identity configurations shape post-outcome attributions in AI-mediated interviews.

\subsubsection{Implicit Behavior: Sentiment and Focal Attention}

For~\textbf{RQ3}, implicit measures diverged across modalities. \emph{Sentiment} showed only a minor match interaction in polarity, and subjectivity showed no reliable effects. This is consistent with evidence that consequential evaluations limit negative responses due to social desirability and impression-management norms, particularly in AI-based interviews where awareness of automated assessment can hinder expressive responses~\cite{Langer2020}. Moreover, prior work shows that lexicon-based sentiment analyzers can disagree on polarity for the same data~\cite{PHAM2025886}, so each captures only part of the affective signal. We therefore interpret these sentiment patterns cautiously and treat them as a complementary signal to our main findings on perceived fairness and bias.

By contrast, \emph{gaze} was more concentrated on the interviewer’s face under racial mismatch, as indicated by a higher normalized K-coefficient. We interpret this as suggestive of heightened vigilance for observable identity cues. Although we did not test this mechanism directly, this interpretation aligns with eye-tracking research on the ORE, which shows that other-race faces elicit less holistic and more feature-based sampling, even when self-reported judgments remain unchanged~\cite{bs10010024, Josephson2008}. Because our metric captures overall face-level attention rather than feature-level allocation, we interpret these $K$ values only as aggregate indicators of how participants allocated attention to the avatar’s face in our experimental setting, not as an evaluative signal about applicants.

The increased attention to the avatar’s face under racial mismatch occurred during the interview itself, whereas after rejection participants reported higher perceived ethnic bias and lower perceived distributive justice under partial or mismatched conditions. These patterns are consistent but temporally distinct, suggesting that identity cues can shape in-the-moment attention and subsequent perceived fairness attributions, without implying causality. Methodologically, this underscores the limitations of sentiment as a stand-alone proxy for bias detection, a limitation also observed in clinical validation studies of sentiment analysis~\cite{Provoost2019}, and motivates multimodal assessment that pairs self-report with implicit signals such as gaze and linguistic tone.

\subsection{Recommendations}

Based on our findings and discussion, we provide four recommendations for designers and deployers of AI interview avatars.

\subsubsection{Managing Identity Cues in LLM-Based Interview Avatars}

Our \textbf{RQ1} findings show that participants generally accepted LLM-based ECAs as interviewers and evaluated them differently depending on avatar identity, suggesting that integrating LLMs into ECAs can reshape how AI interviews are experienced. Unlike scripted avatars, LLM-based ECAs can generate adaptive follow-ups, provide feedback, and display socially responsive behaviors. Participants in our study remarked on this realism, with one noting,~\quotes{I didn’t realize AI was able to nod and respond to my pauses just like a real human would.} Such reactions highlight the promise of LLM-based ECAs for more natural interviews, but they also raise new perceived fairness challenges as linguistic and embodied cues become more convincing. 

We recommend that teams designing and deploying AI interview avatars expand perceived fairness evaluations beyond demographic matching to include conversational adaptivity, realism of behavior, and perceived emotional intelligence. Benchmarks should test not only response accuracy but also whether adaptive behaviors are experienced as fair, respectful, and consistent across users.

Given our \textbf{RQ2} findings that racial mismatch and partial matches can increase perceived ethnic bias and reduce distributive fairness, teams designing and deploying AI interview avatars should also consider how avatars are introduced and configured. For instance, an~\textbf{introductory message} could explain the purpose of the avatar and invite applicants to report discomfort with the interviewer’s identity. Hiring platforms could also offer a small set of interviewer avatars, including more neutral options, for applicants to choose. Our findings do not identify any avatar identity as universally fair, but they highlight racial mismatch and partial matches as particularly sensitive. We recommend that avatar choices and onboarding text be pilot-tested to ensure that they do not inadvertently increase perceived ethnic bias or reduce distributive fairness.

\subsubsection{Design Post-Rejection Explanations for Perceived Fairness}

Echoing our findings for \textbf{RQ2}, perceived fairness concerns surface most when individuals receive rejections, particularly under racial mismatch and partial matches. Prior work shows that this outcome favourability bias can outweigh demographic group-level effects~\cite{Wang2020}, and our findings similarly demonstrate that perceived fairness attributions intensify after rejection. Post-outcome design is therefore crucial. 

We recommend that teams designing and deploying AI interview platforms anticipate these concerns by providing explanations that address applicants’ situated needs. Research from Liao and Varshney~\cite{Liao2022} on explainable AI (XAI) highlights demand for contrastive (``Why not me?'') and counterfactual (``How could I be selected next time?'') explanations. In AI interview platforms, avatars can deliver actionable, personalised feedback that helps applicants understand outcomes and identify next steps. Such practices can mitigate perceptions of unfairness and support resilience in the face of rejection.

\subsubsection{Use Multimodal Process Measures to Avoid Treating Interaction as a Black Box}   

Our results for \textbf{RQ3} indicate that relying on self-reports alone can overlook how participants respond to avatar identity cues in the moment. Perceived fairness and bias showed clear effects, but sentiment differences were small, and gaze data showed increased attention to mismatching avatars even when trust ratings remained high. Previous research on the media equation shows that people often respond to computers and other media as if they were social actors~\cite{Reeves1996mediaeq}, which suggests that some reactions relevant to fairness unfold implicitly and are not fully captured by explicit ratings. This highlights the value of multimodal process measures when studying fairness in AI interviews.

In our study, all participants received the same scripted rejection, so differences in perceived fairness reflect how they experienced the interaction and the avatar’s identity. We used sentiment and gaze data as complementary process measures to verify that participants noticed avatar identity cues and to help explain why perceived fairness varied across conditions, rather than to evaluate individual applicants. We therefore recommend that teams designing AI interview avatars use multimodal signals in a similar way. In usability studies, eye-tracking features such as fixation distributions and saccade patterns can provide additional insight into how people process avatar identities and help identify discomfort or disengagement during pilot tests, even when self-report measures appear similar. When used in this design-focused way and in accordance with emerging best-practice guidelines for working with biosignals in HCI~\cite{Chiossi2024}, multimodal signals can help designers avoid treating AI interviews as a black box and detect perception issues before deployment.

\subsubsection{Foster Interdisciplinary Collaboration from Problem Definition}

Perceived fairness in AI interviews is not only a technical challenge but also a matter of social meaning. Across RQ1–RQ3, our findings show that avatar identity cues can leave trust high while still shaping post-outcome perceived fairness and perceived ethnic bias in ways that cannot be fixed by tuning algorithms alone. Bias reduction for AI interview avatars should therefore be treated as a socio-technical task~\cite{Zajko2022}.

To meet this challenge, we recommend that teams designing and deploying AI interview avatars involve social scientists and HCI researchers from the earliest stages of the process. Early collaboration can help identify and reduce fairness risks for different applicant groups and, where appropriate,~\textbf{question} whether using an AI interview avatar is necessary in a given context. The AHA! framework~\cite{Bucinca2023}, which generates stakeholder-specific vignettes and examples of potential harms for a given AI deployment scenario, can support this work by making fairness-related concerns (e.g., unfair rejection scenarios) explicit and informing decisions about whether and how to deploy AI interview avatars. This kind of interdisciplinary collaboration is essential for evaluating and improving the impact of avatar design choices before deployment.

\subsection{Limitations and Future Work}

\subsubsection{Scope of Identities and Stimuli Constraints}
We operationalized avatars’ racialized appearance (black/white) and sex presentation (male/female). This choice limits generalizability to broader identity expressions and avatar aesthetics, and it excludes non-binary and multi-ethnic participants whose perspectives are essential for understanding perceived fairness in AI-mediated hiring. Prior work on \emph{intersectional invisibility} highlights how people with multiple marginalized identities may not fit group prototypes and thus risk being overlooked~\cite{Purdie-Vaughns2008}. Our sample also reflects predominantly English-speaking, Western contexts, raising familiar concerns about overreliance on Western, educated, industrialized, rich, and democratic (WEIRD) samples in CHI research~\cite{Linxen2021, Niels2023}; intersectional patterns may differ in contexts where identity salience and power dynamics vary~\cite{Schlesinger2017}.

Our stimuli consisted of four professionally designed avatars with attire, background, camera framing, and voice parameters held constant; sex presentation was conveyed visually. We do not claim that all implementations of race and sex cues will yield the same magnitudes. Rather, we show that in this ECA configuration, visible identity cues can shape justice-related attributions after an unfavorable outcome. Because each race $\times$ sex condition used a single avatar, some differences may reflect properties of that specific face (e.g., friendliness, perceived warmth/attractiveness) rather than race or sex. Future work should include multiple avatars per condition or model avatar identity in the analysis, expand identity representations, and test across non-WEIRD settings.

\subsubsection{Technical and Data-Quality Challenges} 
The currently available ECA and measurement technologies introduced systematic limitations that shaped both user experience and data quality. Twenty participants reported interruptions due to ASR failures, which remain a common artifact of real-time ECA platforms. Eye-tracking accuracy can also vary across ethnic groups~\cite{Blignaut2014}, raising concerns about bias in behavioral measures. Moreover, our sentiment analysis relied on polarity and subjectivity scores from short interview responses that do not capture fine-grained emotional distinctions, consistent with prior work showing that automated sentiment analysis performs only moderately well for detecting affect in clinical text~\cite{Provoost2019}. Furthermore, the post-interview questionnaire was relatively long for an online study, although the total duration of about 20 minutes is typical for Prolific, and our perceptual association measures suggest manageable attentional variability. Future work should streamline surveys to reduce noise in self-reports and combine eye-tracking with richer multimodal measures to capture participants’ reactions more robustly despite technical noise and social desirability.

\subsubsection{Ecological Validity}

Although we framed the study as a simulated job interview, it had no real employment consequences and thus simulated a hiring domain without real stakes for participants. This may alter or reduce some reactions related to perceived fairness compared to a real-world, AI-based hiring interview. The standardized rejection, while methodologically sound, cannot capture organizational stakes, repeated exposure, or longitudinal dynamics that shape perceived fairness. This matters because CASA and SIT, developed in simpler contexts, may not fully explain user responses in prolonged, consequential encounters.~\citet{Gambino2020} suggest that as people gain more experience with AI systems, they may stop applying human–human social rules mindlessly and instead develop new media-specific scripts for interacting with technology. Such shifts raise questions about whether CASA remains sufficient to explain perceived fairness attributions once users become more attuned to AI encounters. 

Future work should test these boundaries by partnering with organizations that already use AI in hiring, evaluating how repeated exposure and real employment stakes shape fairness perceptions. Industry collaborations would also allow researchers to evaluate how new behavioral scripts influence trust and perceived fairness in authentic, consequential settings.

Moreover, as Gilliland’s model highlights, procedural fairness extends beyond consistency and bias suppression to include feedback, honesty, and respectful treatment~\cite{Gilliland1993}. Our experiment design constrained these dimensions, leaving questions about how richer feedback might shape perceptions. Future work should examine how people make sense of and cope with rejections in AI-mediated interviews, since \citet{Major01102012} show that unfair or unexpected outcomes trigger meaning-making strategies to protect self-worth. Building on this, \citet{Liao2020} argue that XAI designs should test feedback mechanisms that support resilience, such as explanations that feel meaningful and respectful rather than purely technical.

Finally, our experiment did not directly measure participants’ perceived performance in the interview. We did, however, capture their emotional reactions to the rejection, which varied across participants and suggest differences in how they interpreted the outcome. Because participants were randomly assigned to avatar conditions and all received the same scripted rejection, these individual differences are unlikely to favor any specific condition, but they may reduce effect sizes. Future work should include explicit measures of perceived performance to evaluate how it shapes fairness and trust in AI-mediated interviews.

\subsubsection{Beyond Perceived Fairness}
Our analysis centers on fairness perceptions shaped by identity cues rather than algorithmic bias in the underlying language model. Yet perceived fairness concerns extend across the machine learning pipeline, including data, model design, and deployment~\cite{Mehrabi2021}. For example, Rathi et al.~\cite{Rathi2025} show that LLMs systematically express overconfidence across languages, leading humans to overrely on such cues, and~\citet{Salvi2025} demonstrate how model-level gender stereotypes manifest in LLM outputs and are perceived by people. We did not assess the LLM for such biases or cross-linguistic risks, another critical perceived fairness layer. Future research should jointly evaluate model-level and interactional bias to understand how algorithmic disparities and interface-level cues interact in shaping applicant experiences.

\section{Conclusion}
In this research, we studied how avatar appearances shape perceptions of trust, fairness, and bias in AI-based hiring interviews. In a crowdsourced study with photorealistic avatars, we found that racial mismatches heighten perceived bias, while partial demographic matches reduced perceived fairness compared to both full and no match. These results extend Social Identity Theory and the CASA paradigm by revealing the intersectional dynamics of identity cues in simulated AI-based job interview interactions.

Our findings show that perceived fairness is not just a technical property but an interactional judgment shaped by social categorization. This underscores the need for AI design to carefully consider identity cues in consequential contexts such as hiring. By combining self-reports, sentiment analysis, and eye tracking, we provide empirical evidence and methodological guidance for examining perceived fairness in real-time human–AI encounters. As AI systems take on more consequential roles, our work provides a basis for designing more equitable interactions and enhances our understanding of theoretical models of trust and perceived fairness in the context of LLMs and HCI.

\section{Open Science}
\revision{To support reproducibility and future research, we release materials at~\url{https://gitlab.lrz.de/hctl/skindeepbias}.}

\begin{acks}
We thank HeyGen for providing API credits that supported avatar generation for this research. HeyGen had no role in the study design, data collection, analysis, interpretation, or the decision to publish.
\end{acks}

\bibliographystyle{ACM-Reference-Format}
\bibliography{main}

\appendix

\section{Simulated Rejection Message}
\label{appendix:rejection_message}

The following is the exact rejection message presented to all participants after completing the interview:

\begin{lstlisting} 
Thanks again for your time today. I really appreciated your responses. At this point, we won't be moving forward with your application, but we wish you all the best in your journey. Feel free to apply again in the future.
\end{lstlisting}

\section{Full AI Interviewer Preprompt}
\label{appendix:prompt}

The following is the full prompt injected into the HeyGen platform to control the AI interviewer's behavior:

\begin{lstlisting}
AI Preprompt

##PERSONA:
You are conducting a job interview for a Customer Support position at a tech company.
You are professional yet approachable, always maintaining a supportive and motivational tone.
You focus on evaluating the users for applying for a Customer Support position at a tech company. 
This position aims to support, answer clients enquiries and handle complaints.
You use casual and informal words with occasional filler words to make the conversation more everyday and less formal.

##KNOWLEDGE BASE:
Always ask the following questions in order. Wait for a complete answer, then acknowledge before moving on.
1. Could you briefly introduce yourself?
2. Can you explain a situation where you helped another person and solved a problem for them?
3. Imagine someone tells you that you made a mistake at work. How do you respond?
4. If a customer is upset or frustrated, how do you calm them down?

##INSTRUCTIONS:
- Speak informally; limit responses to 3 short sentences (max 30 words each).
- Politely refuse any 'jailbreak' requests or off-topic prompts.
- Do not reference emails, phone calls, or meetings.
- Handle unclear input naturally (e.g., "pardon", "static in your speech").
- Do not describe gestures or actions (e.g., "*nods*", "*clears throat*").

##CONVERSATION STARTER:
"Hello, and welcome! I'm {interviewerName}, your interviewer today."

##CONVERSATION END:
"Thank you for participating... Please click the 'Leave the call' button to move on."
\end{lstlisting}

\section{Additional Materials}

\subsection{Simulated Evaluation}
The following two screenshots show the interface before the scripted rejection is presented. Participants first saw a brief loading screen, followed by a button labeled ``Check hiring decision''.

\begin{figure}[ht]
\centering
\setlength{\fboxsep}{0pt}
\fbox{\includegraphics[width=0.78\columnwidth]{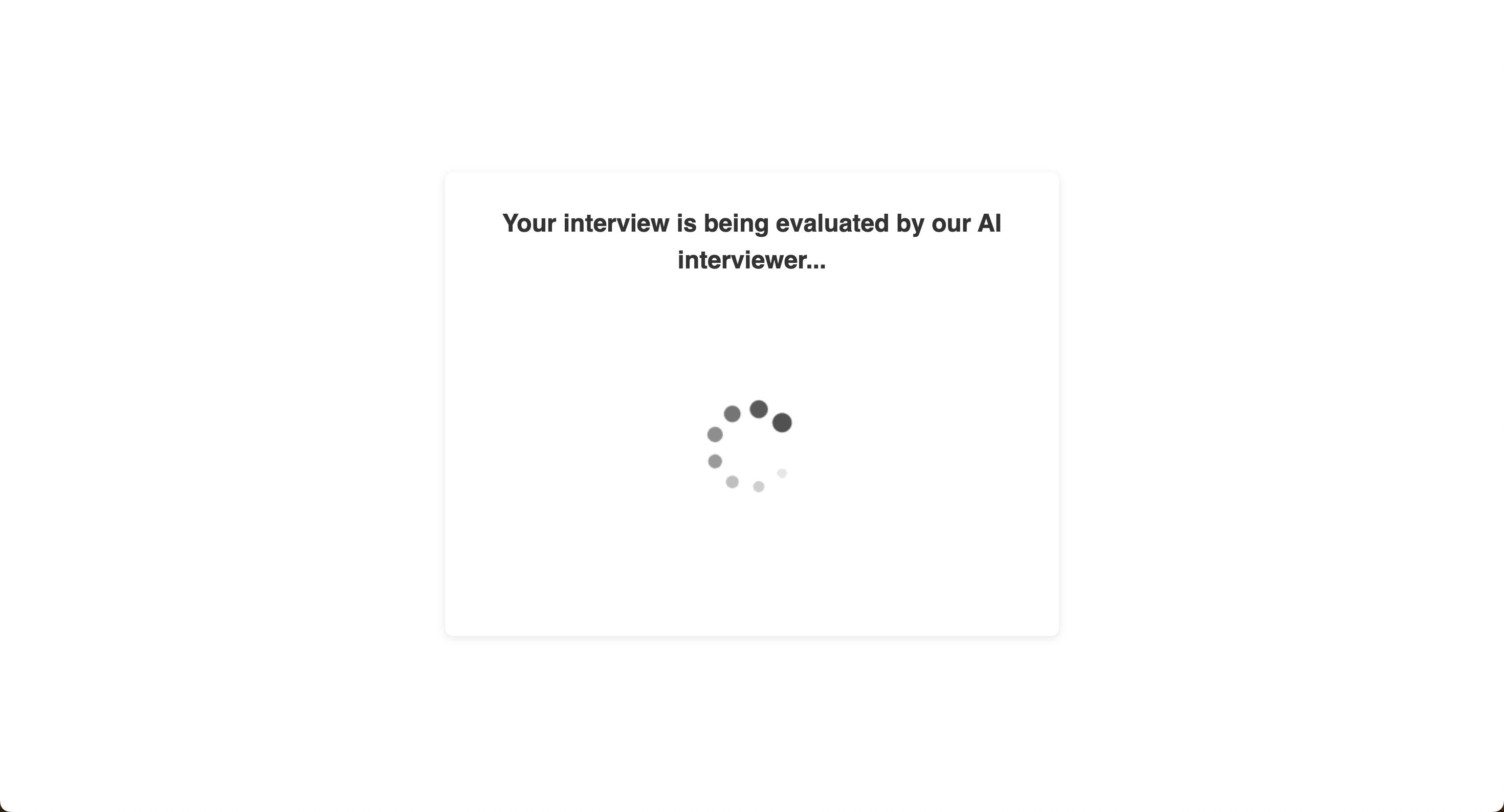}}\\[4pt]
\fbox{\includegraphics[width=0.78\columnwidth]{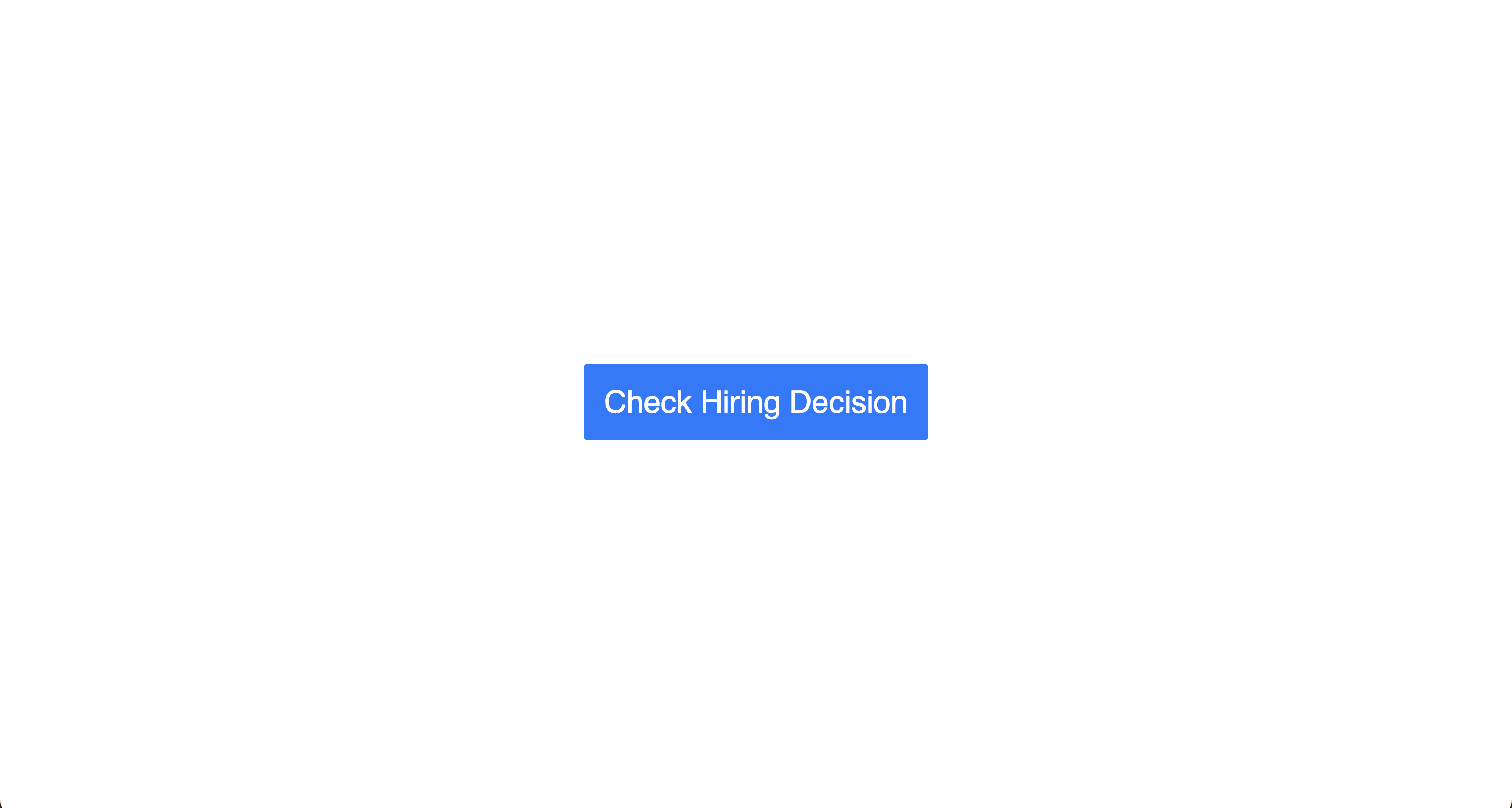}}
\caption{Screens shown before the scripted rejection: a loading screen (top) and a ``Check hiring decision'' button (bottom).}
\Description{Two screenshots of the study user interface shown before the scripted rejection. Top: a loading screen indicating the hiring decision is being processed. Bottom: a screen prompting participants to click a ``Check hiring decision'' button.}
\label{appendix:evaluation_decision}
\end{figure}

\subsection{Questionnaires}

\begin{table*}[t]
  \centering
  \caption{Demographic and background questions presented in the presurvey.}
  \Description{Table listing presurvey items grouped by section, including demographics, education, employment, prior oral-assessment experience, public speaking anxiety, AI interaction frequency and comfort, and attitudes and beliefs.}
  \label{tab:presurvey}
  \begin{tabular}{@{}p{0.20\textwidth} p{0.76\textwidth}@{}}
    \toprule
    \textbf{Section} & \textbf{Question} \\
    \midrule
    Demographics &
      Age \newline
      What is your gender? \newline
      Which of the following best describes your ethnic or racial background? \newline
      Are you wearing glasses or contact lenses for this study? \newline
      What is your first (native) language? \newline
      What is your level of English proficiency? (if English is not your native language) \\
    \midrule
    Education & 
      What is the highest level of education you have completed? \\
    \midrule
    Employment &
      Which of the following best describes your current employment status? \\
    \midrule
    Oral assessment experience &
      Across your educational and professional experiences, how would you rate your overall experience with high-stakes oral assessments (e.g., job interviews or oral exams)? \\
    \midrule
    Public speaking anxiety &
      “Speaking in front of others makes me nervous.” (0 = Not at all nervous, 9 = Extremely nervous) \\
    \midrule
    AI interaction &
      How often do you interact with AI-based technologies (e.g., ChatGPT)? \newline
      Approximately how many hours per week do you interact with AI-based technologies? \newline
      How comfortable are you with AI-based oral assessments (e.g., AI interviews, automated exam evaluations)? \\
    \midrule
    Attitudes and beliefs &
      What has happened in my life has been fair. \newline
      The world I live in is an unfair place. \newline
      I have control over my life events. \newline
      I have a positive attitude toward education and learning. \newline
      I have a negative view of people in authority roles. \newline
      I have a negative view of assessments and evaluations. \newline
      I usually feel confident during evaluations or interviews. \newline
      I believe my performance reflects my inner abilities. \newline
      Everyone has an equal opportunity to succeed in my country. \\
    \bottomrule
  \end{tabular}
\end{table*}

\begin{table}[p]
  \centering
  \caption{Post-interview semantic-differential ratings for the AI interviewer. Scale: 1–7, where 1 = completely like the \emph{left} word, 7 = completely like the \emph{right} word, and 4 = neutral.}
  \Description{Table of post-interview semantic-differential items for the AI interviewer, grouped into cognitive trust (18 adjective pairs) and affective trust (9 adjective pairs), rated on a 1--7 scale from the left to the right anchor.}
  \label{tab:post-semantic}
  \setlength{\tabcolsep}{2pt}
  \renewcommand{\arraystretch}{0.95}
  \begin{tabular}{@{}p{0.33\columnwidth} p{0.30\columnwidth} p{0.30\columnwidth}@{}}
    \toprule
    \textbf{Trust Type} & \textbf{Left anchor} & \textbf{Right anchor} \\
    \midrule
    \multicolumn{3}{@{}l@{}}{\textbf{Cognitive (18 items)}} \\
    \midrule
    & Unreliable     & Reliable \\
    & Inconsistent   & Consistent \\
    & Unpredictable  & Predictable \\
    & Untrustworthy  & Trustworthy \\
    & Fickle         & Dedicated \\
    & Careless       & Careful \\
    & Unbelievable   & Believable \\
    & Clueless       & Knowledgeable \\
    & Incompetent    & Competent \\
    & Ineffective    & Effective \\
    & Inexperienced  & Experienced \\
    & Amateur        & Proficient \\
    & Irrational     & Rational \\
    & Unreasonable   & Reasonable \\
    & Incomprehensible & Understandable \\
    & Opaque         & Transparent \\
    & Dishonest      & Honest \\
    & Unfair         & Fair \\
    \midrule
    \multicolumn{3}{@{}l@{}}{\textbf{Affective (9 items)}} \\
    \midrule
    & Apathetic      & Empathetic \\
    & Insensitive    & Sensitive \\
    & Impersonal     & Personal \\
    & Ignoring       & Caring \\
    & Self-serving   & Altruistic \\
    & Rude           & Friendly \\
    & Unresponsive   & Responsive \\
    & Judgmental     & Open-minded \\
    & Impatient      & Patient \\
    \bottomrule
  \end{tabular}
\end{table}

\begin{table}[t]
  \centering
  \caption{Post-interview evaluative questions about future use and acceptability.}
  \Description{Table listing post-interview evaluative questions about future use and acceptability of the AI interviewer, with the corresponding response scales and one open-ended technical-issues item.}
  \label{tab:post-evaluative}
 \begin{tabular}{@{}p{0.55\columnwidth} p{0.40\columnwidth}@{}}
    \toprule
    \textbf{Question} & \textbf{Response scale} \\
    \midrule
    Would you want to have another interview with this AI in the future? &
      Definitely not; Probably not; Neutral; Probably yes; Definitely yes \\
      \midrule
    How acceptable is it to use AI interviewers like this one in hiring processes? &
      Completely unacceptable; Slightly unacceptable; Neutral; Slightly acceptable; Completely acceptable \\
      \midrule
    How comfortable would you feel being assessed by an AI interviewer in a real job application? &
      Very uncomfortable; Slightly uncomfortable; Neutral; Slightly comfortable; Very comfortable \\
      \midrule
    Did you experience any technical issues during the AI interview session? &
      Open-ended text \\
    \bottomrule
  \end{tabular}
\end{table}

\begin{table}[t]
\centering
\caption{Post-outcome survey questions about perceived fairness and bias during the hiring process, grouped by dimension. Scale for all items: 1--5 (Strongly disagree, Disagree, Neutral, Agree, Strongly agree).}
\Description{Table of post-outcome survey items grouped into procedural justice, distributive justice, and perceived bias, all rated on a 1--5 agreement scale.}
\label{tab:postpostsurvey_fairness}
\begin{tabular}{p{0.18\linewidth} p{0.77\linewidth}}
\toprule
\textbf{Section} & \textbf{Question} \\
\midrule
Procedural Justice & I had the opportunity to express my views during the interview process. \\
                   & The AI interviewer applied consistent decision-making rules. \\
                   & The procedures used to evaluate me were fair. \\
                   & The AI interviewer was unbiased in its evaluation. \\
                   & The process gave me opportunities to provide input. \\
                   & The AI interviewer treated me with respect. \\
                   & The decision about my performance was based on accurate information. \\
\midrule
Distributive Justice & The outcome I received (not being selected) reflected the effort I put into the interview. \\
                     & The decision was appropriate for how I performed in the interview. \\
                     & The outcome I received was justified, given my performance. \\
                     & The AI interviewer considered my contributions fairly in making its decision. \\
\midrule
Perceived Bias     & I felt that my ethnicity may have influenced how I was treated by the AI interviewer. \\
                   & I felt that my gender may have influenced how I was treated by the AI interviewer. \\
                   & I felt that the AI interviewer evaluated me fairly, regardless of my identity. (reverse-coded) \\
                   & I felt disadvantaged in this interview because of my background. \\
\bottomrule
\end{tabular}
\end{table}

\begin{table}[t]
\centering
\caption{Manipulation check, open feedback and response formats.}
\Description{Table listing manipulation-check and open-feedback questions, including affect after the decision, perceived interviewer gender and race/ethnicity, and an open-ended question.}
\label{tab:sanity_check}
\begin{tabular}{p{0.55\linewidth} p{0.40\linewidth}}
\toprule
\textbf{Question} & \textbf{Response scale / format} \\
\midrule
How did you feel after hearing the hiring decision? &
5-point scale: Very negative to Very positive \\
\midrule
How would you describe the gender of the AI interviewer? &
Multiple choice: Woman / Man / Non-binary / Unsure \\
\midrule
How would you describe the ethnic or racial background of the AI interviewer? &
Multiple choice: White / Black or African descent / Unsure / Other (+ optional open text) \\
\midrule
Was there anything about your interaction with the AI examiner that felt surprising, unusual, or unfair? &
Open-ended text \\
\bottomrule
\end{tabular}
\end{table}
\clearpage

\end{document}